\documentclass[universe,review,accept,moreauthors,pdftex]{mdpi}

\usepackage[nolist,nohyperlinks]{acronym}
\usepackage{cleveref}
\usepackage{tikz}
\usetikzlibrary{shapes, arrows, chains, calc, positioning, shapes.geometric, arrows.meta}
\usepackage{subcaption}
\usepackage{ amssymb }
\usepackage{placeins}

\newcommand{\uv}{\ensuremath{(u, v)}}

\firstpage{1} 
\articlenumber{accepted for publication}
\doinum{10.3390/------}
\pubvolume{VLBI Science Applications}
\pubyear{2022}
\copyrightyear{2022}
\externaleditor{Academic Editor: }
\history{Received: date; Accepted: date; Published: date}

\Title{Software and techniques for VLBI data processing and analysis}

\Author{Michael Janssen$^{1}$*\orcidA{}, Jack~F.~Radcliffe$^{2,3,4}$\orcidC{}, Jan Wagner$^{1}$\orcidB{}}

\AuthorNames{Michael Janssen, Jack F. Radcliffe, Jan Wagner}

\address{%
$^{1}$ \quad Max-Planck-Institut für Radioastronomie, Bonn, Germany\\
$^{2}$ \quad Department of Physics, University of Pretoria, Lynnwood Road, Hatfield, Pretoria 0083, South Africa\\
$^{3}$ \quad Jodrell Bank Centre for Astrophysics, University of Manchester, Oxford Road, Manchester M13 9PL, UK \\
$^{4}$ \quad National Institute for Theoretical and Computational Sciences (NITheCS) South Africa}

\corres{Correspondence: mjanssen@mpifr-bonn.mpg.de; Tel.: +49-228-525-431}

\abstract{Very-long-baseline interferometry (VLBI) is a challenging observational technique, which requires in-depth knowledge about radio telescope instrumentation, interferometry, and the handling of noisy data. The reduction of the raw data is mostly left to the scientists and demands the use of complex algorithms implemented in comprehensive software packages. The correct application of these algorithms necessitates a good understanding of the underlying techniques and physics that are at play. The verification of the processed data produced by the algorithms demands a thorough understanding of the underlying interferometric VLBI measurements. This review describes the latest techniques and algorithms that scientists should know about when analyzing VLBI data.}


\begin{acronym}
\acro{sn}[$S\,/\,N$]{signal-to-noise ratio}
\acro{vlbi}[VLBI]{very-long-baseline interferometry}
\acroplural{vlbi}[VLBI]{Very long baseline interferometry}
\acro{grmhd}[GRMHD]{general relativistic magnetohydrodynamics}
\acro{mhd}[MHD]{magnetohydrodynamics}
\acro{as}[as]{arcseconds}
\acro{agn}[AGN]{Active Galactic Nuclei}
\acro{llagn}[LLAGN]{low-luminosity AGN}
\acro{cena}[Cen~A]{Centaurus A}
\acro{sgra}[Sgr\,A*]{Sagittarius~A*}
\acro{agn}[AGN]{active galactic nuclei}
\acro{eht}[EHT]{Event Horizon Telescope}
\acro{gmva}[GMVA]{Global mm-VLBI Array}
\acro{bhc}[BHC]{\href{https://blackholecam.org}{BlackHoleCam}}
\acro{tanami}[TANAMI]{Tracking Active Galactic Nuclei with Austral Milliarcsecond Interferometry}
\acro{smbh}[SMBH]{supermassive black hole}
\acroplural{smbh}[SMBHs]{supermassive black holes}
\acro{aa}[ALMA]{Atacama Large Millimeter/submillimeter Array}
\acro{ap}[APEX]{Atacama Pathfinder Experiment}
\acro{pv}[PV]{IRAM~30\,m Telescope}
\acro{jc}[JCMT]{James Clerk Maxwell Telescope}
\acro{lm}[LMT]{Large Millimeter Telescope Alfonso Serrano}
\acro{sp}[SPT]{South Pole Telescope}
\acro{sm}[SMA]{Submillimeter Array}
\acro{az}[SMT]{Submillimeter Telescope}
\acro{chandra}[Chandra]{Chandra X-ray Observatory}
\acro{c}[c]{speed of light}
\acro{pc}[pc]{parsec}
\acro{gr}[GR]{general relativity}
\acro{aips}[\texttt{AIPS}]{\href{http://www.aips.nrao.edu}{\texttt{Astronomical Image Processing System}}}
\acro{casa}[\texttt{CASA}]{\href{https://casa.nrao.edu}{\texttt{Common Astronomy Software Applications}}}
\acro{symba}[\texttt{SYMBA}]{\href{https://bitbucket.org/M_Janssen/symba}{\texttt{SYnthetic Measurement creator for long Baseline Arrays}}}
\acro{rpicard}[\texttt{rPICARD}]{\href{https://bitbucket.org/M_Janssen/picard}{\texttt{Radboud PIpeline for the Calibration of high Angular Resolution Data}}}
\acro{hops}[\texttt{HOPS}]{\href{https://www.haystack.mit.edu/tech/vlbi/hops.html}{\texttt{Haystack Observatory Postprocessing System}}}
\acro{hbt}[HBT]{Hanbury Brown and Twiss}
\acro{tov}[TOV]{Tolman-Oppenheimer-Volkoff}
\acro{mri}[MRI]{magnetorotational instability}
\acro{adaf}[ADAF]{advection-dominated accretion flow}
\acro{adios}[ADIOS]{adiabatic inflow-outflow solution}
\acro{cdaf}[CDAF]{convection-dominated accretion flow}
\acro{bz}[BZ]{Blandford-Znajek}
\acro{bp}[BP]{Blandford-Payne}
\acro{em}[EM]{electromagnetic}
\acro{blr}[BLR]{broad-line region}
\acro{nlr}[NLR]{narrow-line region}
\acro{ism}[ISM]{interstellar medium}
\acro{edf}[eDF]{electron distribution function}
\acro{pic}[PIC]{particle-in-cell}
\acro{sane}[SANE]{standard and normal evolution}
\acro{mad}[MAD]{magnetically arrested disk}
\acro{jive}[JIVE]{\href{http://www.jive.eu}{Joint Institute for VLBI ERIC}}
\acro{nrao}[NRAO]{\href{https://www.nrao.edu}{National Radio Astronomy Observatory}}
\acro{muas}[$\mu$as]{microarcseconds}
\acroplural{agn}[AGN]{Active galactic nuclei}
\acro{jy}[Jy]{Jansky}
\acro{pa}[PA]{position angle}
\acro{srmhd}[SRMHD]{special relativistic magnetohydrodynamics}
\acro{fov}[FOV]{field of view}
\acroplural{fov}[FOVs]{fields of view}
\acro{sefd}[SEFD]{system equivalent flux density}
\acroplural{sefd}[SEFDs]{system equivalent flux densities}
\acro{tec}[TEC]{total electron content}
\acro{eop}[EOP]{IERS Earth Orientation Parameter}
\acroplural{eop}[EOPs]{IERS Earth Orientation Parameters}
\acro{ms}[\texttt{MS}]{\texttt{Measurement Set}}
\acro{rfi}[RFI]{radio frequency interference}
\end{acronym}
\begin{document}



\section{Introduction}

In recent years, several novel techniques and methods have been developed for the processing and analysis of \ac{vlbi} data, related primarily to the new needs and capabilities of novel \ac{vlbi} instruments. For instance, the growing number of telescopes participating in large and global alliances of \ac{vlbi} arrays \citep[e.g.,][]{2011saip.conf..473G, 2016mks..confE..17D, 2018PASP..130a5002M, 2020arXiv200702347V, 2021AdSpR..68.3064B}, significant increases in bandwidths via improved gigabit per second (Gbps) recording rates \citep[e.g.,][]{2013PASP..125..196W,2015PASP..127.1226V}, multi-frequency receiver capabilities \citep[e.g.,][]{2008IJIMW..29...69H}, advancements towards observing increasingly shorter wavelength emission to gain resolving power and counter self-absorption effects in the observed sources \citep[e.g.,][]{2017A&ARv..25....4B,EHT2019I}, and prospects of the future next-generation Very Large Array \citep[ngVLA,][]{2018ASPC..517...15S,2018ASPC..517....3M}, \ac{vlbi} capabilities of the Square Kilometre Array \citep[SKA,][]{2009IEEEP..97.1482D,2015aska.confE.143P}, next-generation Event Horizon Telescope \citep[ngEHT,][]{2019BAAS...51g.256D}, as well as millimeter space VLBI observatories \citep[e.g.,][]{2014PhyU...57.1199K, 2019BAAS...51g.235J, 2020AdSpR..65..821F, 2021ExA....51..559G}.

Lately, the localization of a repeating fast radio burst \citep{2017ApJ...834L...8M} and black hole images \citep{EHT2019I, eht-sgra-PaperI} have been continuum total intensity \ac{vlbi} science highlights. More complex \ac{vlbi} methods, which require a few additional steps, enable the study of magnetic field signatures through polarized synchrotron radiation, spectral line emission from astronomical masers, and wide \acp{fov} combined with the \ac{vlbi} resolving power.

With polarimetric \ac{vlbi} \citep[e.g.,][]{2018Galax...7....5G,2020AdSpR..65..731G,2021ApJ...910L..12E, 2021ApJ...910L..13E, 2021Galax...9...58G}, the role that magnetic fields play in the acceleration, collimation, and morphology of \ac{agn} jets can be studied across a wide range of scales. At the smallest scales, parameters of the central object responsible for producing the jet can be inferred to test for the presence of an event horizon, measure the spin of the compact object, and assess the role of the magnetic field in the jet launching.


Through spectral line \ac{vlbi} \citep[e.g.,][]{1996A&ARv...7...97H, 2003Momjian, 2005A&A...432..161G, 2006ApJ...645..337H, 2009ApJ...695..287R, 2011AN....332..461B, 2011A&A...535L...8G, 2013ApJ...767..154R, 2017A&A...597A..43G, 2019A&A...631A..74M}, maser emission from star forming regions, H I absorption, ejected circumstellar envelopes from giant stars, as well as gas in the vicinity of \ac{agn} that may be part of inflows or outflows can be mapped. Through Doppler shifts, the kinematics and physical conditions of the sources can be studied. Furthermore, angular-diameter distances to megamaser disk galaxies can be used for cosmology.

Wide-field \ac{vlbi} \citep[e.g.,][]{w1,w2,w3,w4,w5,2019ApJ...875..100D} enables powerful astrometric studies, the mapping of distances to pulsars in our galaxy, and the search for gravitationally lensed radio sources.
Moreover, faint sources in the \ac{fov} can be detected, details in large extended structures such as \ac{agn} jet hotspots can be imaged when the instrument's sensitivity is sufficient, and star formation radio emission can be distinguished from \ac{agn} activity in deep field studies.

Additionally, \ac{vlbi} observations of spacecraft should be mentioned \citep[e.g.,][and references therein]{2012A&A...541A..43D}.
With a sufficiently stable and strong onboard radio emitter, spacecraft can be tracked with high accuracy, allowing for a wide range of planetary science applications, gravimetric studies, and tests of general relativity for example.

In this work, we review the typical sequence of \ac{vlbi} processing steps and highlight the latest established open-source software packages and techniques.
This information is targeted primarily at scientists who have obtained data from a \ac{vlbi} observation and want to get an overview of the available tools for data processing and scientific analyses.
We focus on the latest state-of-the-art algorithms and do not give a historical overview.
We also do not discuss the scheduling of \ac{vlbi} observations, which principal investigators (PIs) should do together with observatories and the help of the \texttt{SCHED} software.\footnote{\url{http://www.aoc.nrao.edu/software/sched}. See also \url{https://github.com/jive-vlbi/sched} and \url{https://planobs.jive.eu}.}
For a broad theoretical \ac{vlbi} background, the reader is referred to \citet{TMS}. Definitions of technical terms used in this work that new \ac{vlbi} users might be unfamiliar with are given in \Cref{tab:conventions}.

\ac{vlbi} measurements with the long baselines of the Low Frequency Array (LOFAR) are also possible, but require unique calibration strategies that are beyond the scope of this work. For example, due to the prevalence of direction-dependent effects, very low observing frequencies, and large data volumes. For details, the reader is referred to \citet{lofarvlbi}.

This review is structured as follows: \Cref{sec:corr} describes the correlation of \ac{vlbi} data.
\Cref{sec:signal} details the ``signal stabilization'', which encompasses all data calibration steps needed before the data can be averaged in time and frequency.
\Cref{sec:flux} outlines the amplitude gain corrections for telescopes that are needed for a flux density calibration in physical units.
\Cref{sec:image} explains how the sky brightness distribution of the observed source can be reconstructed from the calibrated data.
\Cref{sec:special} addresses how special science cases of polarization, spectral line, and wide-field \ac{vlbi} are to be handled.
Use-cases and algorithms for the synthetic generation of \ac{vlbi} data are given in \Cref{sec:synth}.
A summary is presented in \Cref{sec:summary}. 

\section*{List of Abbreviations}

\begin{tabbing}
    AGN \hspace{1cm} \= Active galactic nuclei\\
    ALMA \> Atacama Large Millimeter/submillimeter Array\\
    ASCII \> American Standard Code for Information Interchange\\
    ASIC \> Application-specific integrated circuit\\
    CLEAN \> Imaging algorithm for an incomplete Fourier coverage\\
    EHT \> Event Horizon Telescope\\
    EVN \> European VLBI Network \\
    EVPA \> Electric vector polarization angle\\
    EOP \> Earth orientation parameter\\
    FFT \> Fast Fourier transform\\
    FPGA \> Field-programmable gate array\\
    FOV \> Field of view\\
    FPT \> Frequency-phase-transfer\\
    Gbps \> Gigabit per second\\
    GMVA \> Global mm-VLBI array\\
    GPS \> Global Positioning System\\
    GPU \> Graphics processing unit\\
    GRMHD \> General relativistic magnetohydrodynamics\\
    HPC \> High-performance computing\\
    IDG \> Image-domain gridder\\
    IERS \> International Earth Rotation and Reference Systems Service\\
    JIVE \> Joint Institute for Very Long Baseline Interferometry European Research Infrastructure\\ \> Consortium \\
    KVN \> Korean VLBI Network\\
    LCP \> Left circularly polarized\\
    LOFAR \> Low Frequency Array\\
    MAD \> Magnetically arrested accretion disc\\
    MERLIN \> Multi-Element Radio Linked Interferometer Network\\
    MFS \> Multi-frequency synthesis\\
    MPI \> Message Passing Interface\\
    MSC \>  Multi-scale CLEAN\\
    MSSC \> Multi-source self-calibration\\
    MS-MFS \> Multi-scale multi-frequency synthesis (combines MSC with MT-MFS)\\
    MT-MFS \> Multi-term multi-frequency synthesis\\
    ngEHT \> Next-generation Event Horizon Telescope\\
    ngVLA \> Next-generation Very Large Array\\
    NRAO \> National Radio Astronomy Observatory\\
    RCP \> Right circularly polarized\\
    RML \> Regularized maximum likelihood\\
    PI \> Principal investigator\\
    PSF \> Point spread function\\
    RF \> Radio frequency\\
    RFI \> Radio frequency interference\\
    SANE \> Standard and normal evolution accretion state\\
    SEFD \> System equivalent flux density\\
    SFPR \> Source frequency phase referencing\\
    SKA \> Square Kilometre Array\\
    $S\,/\,N$ \> Signal-to-noise ratio\\
    TCP \> Transmission Control Protocol\\
    TEC \> Total electron content\\
    UDP \> User Datagram Protocol\\
    UT \> Universal Time\\
    VERA \> VLBI Exploration of Radio Astrometry\\
    VGOS \> VLBI Global Observing System\\
    VLA \> Very Large Array\\
    VLBA \> Very Long Baseline Array\\
    VLBI \> Very-long-baseline interferometry
\end{tabbing}

\begin{table}[h!]
\caption{Basic terms, definitions, and concepts are listed here, which new \ac{vlbi} users might be unfamiliar with or which might have slightly different meanings in other related works.}
\centering
\begin{tabular}{p{1.7cm}p{13.5cm}}
\toprule
\textbf{Name} & \textbf{Meaning}\\
\midrule
Telescope frontend & Equipment ``directly attached to the front of a telescope'', used to detect and encode the sky signal. Receivers (e.g., bolometers) are frontend equipment.\\ \\
Telescope backend & Equipment at the telescope that is used to process and store the signal from the frontend. Block Downconverters, which mix the sky signal down to a lower frequency range (heterodyning), analog-to-digital converters, which digitize the data, and data recorders, which store the digitized measurements on hard drives, are backend equipment.\\ \\
Baseband data		& The recorded data at a telescope that will be used for the correlation. More precisely, the filtered, down-converted, sampled, and quantized electric field measurements stored in the backends.\\ \\
Signal stabilization & Described in \Cref{sec:signal}: The collection of all post-correlation calibration measures, excluding the a priori flux density calibration (\Cref{sec:flux}). The signal stabilization is often referred to as fringe-fitting, but it also involves additional steps, e.g., corrections for bandpass responses and corrections for atmospheric phase turbulence.\\\\
Delay		& Residual post-correlation phase-slopes as a function of frequency (e.g., due to atmospheric path length differences). To be corrected in the signal stabilization step.\\ \\
Rate		&  Residual post-correlation phase-slopes as a function of time (e.g., due to Doppler shifts) To be corrected in the signal stabilization step.\\\\
Fringe-fit FFT & The fast Fourier transform step of common fringe-fitting algorithms. Transforms the visibilities from time, frequency space into a rate, delay space, where the peaks are to be found. The height of the peaks corresponds to the strength of the signal.\\\\
Low $\nu$ & An observing frequency below 20\,GHz.\\\\
High $\nu$ & An observing frequency above 20\,GHz.\\\\
Allan deviation & A measure of frequency stability \citep{1966Allan}. An easy-to-follow derivation and description of the Allan deviation equation is given in Section 9.5.1 of \citet{TMS}.\\\\
\texttt{VEX} file & ``VLBI EXperiment'' file, which describes the VLBI setups and observing schedules in a standardized text format (\url{https://vlbi.org/wp-content/uploads/2019/03/vex-definition-15b1.pdf}).\\\\
\texttt{FITS-IDI}\; file & ``FITS Interferometry Data Interchange Convention'' standardized file format for visibility and VLBI metadata built upon the upon the standard \texttt{FITS} format (\url{https://fits.gsfc.nasa.gov/registry/fitsidi/AIPSMEM114.PDF}).\\\\
\texttt{ANTAB} file & ``Antenna table'', which contains station gain and system temperature information in a simple text file format (\url{http://www.aips.nrao.edu/cgi-bin/ZXHLP2.PL?ANTAB}).\\
\bottomrule
\label{tab:conventions}
\end{tabular}
\end{table}

\section{Correlation}
\label{sec:corr}

PIs are usually provided with correlated data, but it is instructive for scientists to understand the fundamentals of interferometric observations and upstream processing applied to their data.


There are two types of astronomical interferometers: Connected-element interferometers, and very long baseline interferometric networks that are used in astronomical and geodetic VLBI observations. Connected-element arrays can be ``phased'', producing beamforming and phased-sum time domain signal output (e.g., \cite{2016JAI.....541006P,2018PASP..130a5002M}). Such phased arrays can participate in VLBI, and are equivalent to a single large dish with a narrow beam.

For interferometry, the recorded baseband data (\Cref{tab:conventions}) of all antennas are combined at a ``correlator'' computing system. 
In a connected-element interferometer, telescopes are connected over short (walkable) distances to a central correlator via RF-over-fiber or high speed network links, and share a common time and frequency standard. The short distances within an array permit very wide baseband signals (currently up to 64\,GHz of bandwidth) from a large number of antennas (currently up to 10,000) to be transferred and processed at the array correlator in real time. The correlator usually employs hardware for the signal processing (FPGA, ASIC) due to power efficiency and operating cost considerations.
With real-time correlation, the voluminous (often $>$ 10~Tbit/s aggregate) baseband data does not need to be stored. Only the final time averaged data products are kept.

\ac{vlbi} networks consist of relatively few ($\lesssim$ 20) telescopes and, unlike in most connected-element arrays, they are often heterogeneous with different frequency tuning constraints, digital baseband bandwidths, baseband data formats, and no shared time and frequency standard. The large distances between telescopes located across the globe and in orbits around the Earth for space \ac{vlbi} make real-time data transfer challenging.

Although there are recent VLBI hardware correlators, e.g., the 16-station x 8~Gbps Korean VLBI Network \citep[KVN,][]{KVN1,KVN2,KVN3} and VLBI Exploration of Radio Astrometry \citep[VERA,][]{VERA} correlator at the Korea-Japan Correlation Center \citep{2015JKAS...48..125L}, software correlators are currently more common. The flexible signal processing in software can better address possible complications that arise from the heterogenous set of telescopes. In addition, new observing modes are more rapidly implemented in software correlators.

\subsection{Software Correlators}


Two typical open source software VLBI correlators are \texttt{DiFX}\footnote{\texttt{DiFX}: \url{https://www.atnf.csiro.au/vlbi/dokuwiki/doku.php/difx/documentation}.} \cite{2007PASP..119..318D,2011PASP..123..275D} and \texttt{SFXC}\footnote{\texttt{SFXC}: \url{https://www.jive.eu/jivewiki/doku.php?id=sfxc}.} \cite{2015ExA....39..259K}. \texttt{DiFX} has a wide developer community and user base including  the \ac{nrao}/Very Long Baseline Array \citep[VLBA,][]{VLBA}, the Event Horizon Telescope \citep[EHT,][]{2019ApJ...875L...2E}, and geodetic services. \texttt{SFXC} is more native to the European VLBI Network \citep[EVN, see][and references therein]{2010evn..confE..11P} environment.
%



Being ``embarrassingly parallel'', the correlation process is highly suited for cluster computing. Recently, graphics processing units (GPUs) are started to be used. One production GPU correlator is the 6-station x 16~Gbps near-realtime {\em RasFX} of the Quasar VLBI network \cite{surkis2017rasfxvgos,SHUYGINA2019150}. Other GPU-based VLBI correlators are under development or in a demonstrator stage, including an Extended-KVN GPU correlator and a possible correlator for VLBI space science \cite{Zhang_2021}. A GPU-accelerated version of \texttt{DiFX} is under development as well.\footnote{\url{https://adacs.org.au/project/gpu-acceleration-of-the-difx-software-correlator}.}


\subsection{Station Clock Model}\label{s:tftransfer}


Typically, each \ac{vlbi} telescope operates its own frequency standard, with an Allan deviation \citep[][\Cref{tab:conventions}]{1966Allan} of $<10^{-13}$\,s/1\,s and $<10^{-13}$\,s/100\,s to surpass atmospheric turbulence \cite{2012AJ....144..121R,Clivati_2020}. Receivers and backends are locked to the reference, and backends time-tag their digital baseband output data. An absolute time reference is provided by GPS, with an uncertainty of around $\pm 20$\,ns due to ionospheric diurnal variation. Possibilities of sharing clock signals over long distances are currently being investigated \citep[e.g.,][]{2017A&A...603A..48K,Clivati_2020}.

Without a shared clock signal, time-frequency transfer is emulated by a clock model that is applied at the correlator. The model consists of station clock offsets and drifts over time of the station frequency standards. Clock offsets are the differences measured between the backend data time-tag and the absolute Universal Time (UT). Drifts are measured via a continuous comparison against the GPS absolute time, usually fit to first order (linear drift) or sometimes to a higher order (acceleration) in space VLBI \cite{2017JAI.....650004L}. 

Software correlators tend to be more flexible with their clock models than their hardware counterparts, handling offsets of several seconds at sub-microsecond granularity, and drifts of $10^{-9}$\,s/1\,s or larger.

\subsection{Processing at the Correlator}

The input for the correlator data processing are the baseband signals $E_r(t)$ from all telescopes. Appendix~\ref{sec:corr-data} describes how the recorded data are shipped to the correlation facility. Processing involves trivial computations at a high data rate. It produces time averaged mutual coherences between pairs of telescopes, analytically given by 
\begin{equation}
    \Gamma_{r_1,r_2}{(u,v,\tau)} = \lim\limits_{T \to \infty} \frac{1}{2T} \int_{-T}^{T} E_{r_1}(t) E^*_{r_2}(t-\tau) \mathrm{d}t\;,
\end{equation}
at $\tau\!=\!0$, over an integration time~$T$, for telescopes at locations $r_1$ and $r_2$; see also Equation~15.1 in \cite{TMS}. Coherences between all telescope pairs and across their different spatial separations, or \uv{}-plane spatial frequencies, are called interferometric ``visibilities'' (cf. \S~4.1 of \cite{TMS}).

According to the {\em van Cittert-Zernike theorem}, these visibilities are the Fourier counterpart of the angular source brightness distribution, $I(l,m)$, related through
\begin{equation}
    \Gamma_{r_1,r_2}(u, v, 0) = \iint I(l,m) \exp\left[-2\pi i (ul+vm)\right] \mathrm{d}l \mathrm{d}m\;;
\end{equation}
see Equations~3.7 and 15.7 in \cite{TMS}. Here, the line-of-sight directions across the brightness distribution of the sky source are given by the unit vectors $(l, m, \sqrt{1-l^2-m^2})$. Visibility data produced by the correlator can thus be used for Fourier imaging -- after a series of calibration steps described in later sections.

In practice the correlator determines $\Gamma_{r_1,r_2}(u, v, n\Delta\tau)$ over a discrete range of time ``delay lags'', $n\Delta\tau$, or equivalently in the spectral domain, spectral channels $N\Delta\nu$ across the baseband. 

\subsection{Correlator Delay Model}

The correlator attempts to phase-align stations for maximum coherence to occur around $\tau=0$. This requires a delay model that includes the station clock models (\Cref{s:tftransfer}), and several predicted geometric and propagation-path-related delays. These stem from observing geometry, Earth orientation, tidal/loading effects on station coordinates, tropospheric and ionospheric propagation, near-field wavefront and relativistic aberration effects, telescope orbit and orientation parameters for space VLBI, etc. The delay model can be produced with the common NASA GSFC \texttt{Calc/Solve} package or software specific to space VLBI or near objects \citep{Gordon2004calc,2006JGeod..80..137S,2017JAI.....650004L}.


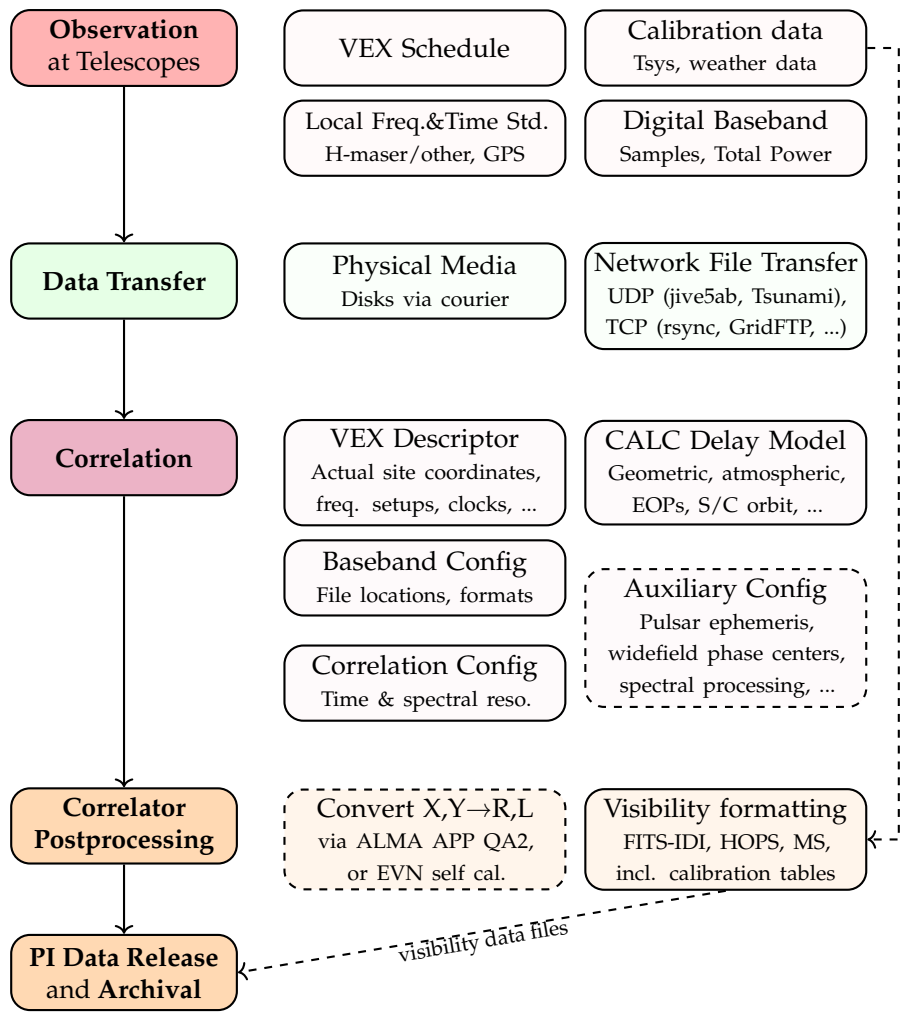
\begin{figure}[H]
\centering
\begin{tikzpicture}
  \def \yspc {-2cm}
  \def \yitemspc {-0.20cm}
  \def \yitemspcB {-0.40cm}
  \def \xpsc {3cm}
  \tikzstyle{block} = [
    draw=black,
    minimum width = 3.0cm, minimum height = 1.0cm,
    align = center, thick, rounded corners=.25cm
    ]
  \tikzstyle{blockitem} = [
    draw=black,
    minimum width = 3.0cm, minimum height = 1.0cm,
    align = center, thick, rounded corners=.25cm,
    text width=3.5cm
    ]
  \tikzstyle{blockoptionalitem} = [
    draw=black,dashed,
    minimum width = 3.0cm, minimum height = 1.0cm,
    align = center, thick, rounded corners=.25cm,
    text width=3.5cm
    ]
    \tikzstyle{arrow} = [thick, ->]

  \node[block, fill=red!30] (P1) {\bfseries Observation\\at Telescopes};
  \node[blockitem, fill=red!2, right of=P1, xshift=\xpsc] (P1a) {VEX Schedule};
  \node[blockitem, fill=red!2, right of=P1a, xshift=\xpsc] (P1b) {Calibration data\\{\footnotesize Tsys, weather data}};
  \node[blockitem, fill=red!2, below of=P1a, yshift=\yitemspc] (P1c) {\small Local Freq.\&Time Std.\\{\footnotesize H-maser/other, GPS}};
  \node[blockitem, fill=red!2, right of=P1c, xshift=\xpsc] (P1d) {Digital Baseband\\{\footnotesize Samples, Total Power}};

  \node[block, fill=green!10, below of=P1, yshift=-2.1cm] (P2) {\bfseries Data Transfer};
  \node[blockitem, fill=green!2, right of=P2, xshift=\xpsc] (P2a) {Physical Media\\{\footnotesize Disks via courier}};
  \node[blockitem, fill=green!2, right of=P2a, xshift=\xpsc, yshift=-0.2cm] (P2b) {Network File Transfer\\{\footnotesize UDP (jive5ab, Tsunami), TCP (rsync, GridFTP, ...)}};
  \draw[arrow] (P1) -- (P2);

  \node[block, fill=purple!30, below of=P2, yshift=-1.35cm] (P3) {\bfseries Correlation};
  \node[blockitem, fill=purple!2, right of=P3, xshift=\xpsc, yshift=-0.2cm] (P3a) {VEX Descriptor\\{\footnotesize Actual site coordinates, freq. setups, clocks, ...}};
  \node[blockitem, fill=purple!2, right of=P3a, xshift=\xpsc] (P3b) {CALC Delay Model\\{\footnotesize Geometric, atmospheric, EOPs, S/C orbit, ...}};
  \node[blockitem, fill=purple!2, below of=P3a, yshift=\yitemspcB] (P3c) {Baseband Config\\{\footnotesize File locations, formats}};
  \node[blockoptionalitem, fill=purple!2, below of=P3b, yshift=-1.20cm] (P3d) {Auxiliary  Config\\{\footnotesize Pulsar ephemeris, widefield phase centers, spectral processing, ...}};
  \node[blockitem, fill=purple!2, below of=P3c, yshift=\yitemspcB] (P3e) {Correlation Config\\{\footnotesize Time \& spectral reso.}};
   \draw[arrow] (P2) -- (P3);
  
  \node[block, fill=orange!30, below of=P3, yshift=-3.9cm] (P4) {\bfseries Correlator\\\bfseries Postprocessing};
  \node[blockoptionalitem, fill=orange!8, right of=P4, xshift=\xpsc, yshift=-0.18cm] (P4a) {Convert X,Y\textrightarrow R,L\\{\footnotesize via ALMA APP QA2, or EVN self cal.}};
  \node[blockitem, fill=orange!8, right of=P4a, xshift=\xpsc] (P4b) {Visibility formatting
  \\{\footnotesize FITS-IDI, HOPS, MS, incl. calibration tables}};
  \draw[arrow] (P3) -- (P4);
  \draw[arrow,dashed] (P1b.east) -| (10.3,0) |- (P4b.east);
  
  \node[block, fill=orange!30, below of=P4, yshift=-0.95cm] (P5) {\bfseries PI Data Release\\and \bfseries Archival};
  \draw[arrow] (P4) -- (P5);
 
  \draw[arrow,dashed] (P4b.south) -- (P5.east) node[midway,sloped,yshift=-0.15cm] {\footnotesize visibility data files};

\end{tikzpicture}
\caption{VLBI observation and data processing. Per-telescope baseband and calibration data are collected during VLBI observations. Calibration data are bundled with the final visibility data release to the PI. Delay models for far-field VLBI are commonly derived via NASA GSFC \texttt{CALC}~9/10, for near-field VLBI via {\em c5++} or others. Dashed boxes indicate optional processing, such as, pulsar processing, space VLBI antenna support, multi phase center for wide-field VLBI, spectral windowing (in \texttt{SFXC}) or custom channelization (zoom bands in \texttt{DiFX}). Correlation itself can run on hardware ranging from Raspberry PI, HPC clusters, to Cloud platforms.}\label{fig:correlatorFlow}
\end{figure}


The a~priori delay model cannot predict all encountered delay components. Visibility data correlated with a wide enough delay range $n\Delta\tau$ and at sufficiently high temporal cadence allow unmodeled delay residuals, coming from random atmospheric turbulence for example, to be calibrated out via fringe-fitting (\Cref{tab:conventions}) during post-processing (\Cref{sec:signal}).

\subsection{Correlator Parameters}
\label{sec:corrparams}

Software correlators such as \texttt{DiFX} and \texttt{SFXC} are highly configurable and support various science and observing modes. Basic parameters are the PI-requested integration time and spectral resolution. 
Several additional data and settings flow into the correlator configuration, as illustrated in \Cref{fig:correlatorFlow}.

In general, configuration files include a \texttt{VEX} file\footnote{\texttt{VEX}: \url{https://vlbi.org/wp-content/uploads/2019/03/vex-definition-15b1.pdf}.} derived from the observing schedule, edited as necessary to describe the actual rather than nominal frequency setups and baseband data layouts each station used during the observation. 
%

Depending on correlator features, additional parameters may include, e.g., 1) pulsar ephemeris data and binning/gating setups, 2) additional interferometric phase centers for wide-field VLBI within the telescopes' primary beam, 3) settings for baseband spectral slicing or concatenation via \texttt{DiFX} ``zoom'' and ``output'' bands. These special band modes allow uniform-bandwidth visibility data records to be produced even for VLBI experiments where telescopes had incompatible frequency setups. 4) \texttt{DiFX} can also extract phase information of phase/pulse cal tone combs in the baseband data, if tones were injected during the observation. These combs are used for instrumental delay/phase calibration purposes, especially in geodetic VLBI. (5) At the VLBA, \texttt{DiFX} can extract on/off digital power levels of a synchronously switched calibration signal and produce continuous system temperature measurements (\Cref{sec:flux}). (6) \texttt{SFXC} offers data windowing or weighted-overlap-add processing, which shape the frequency response of the spectral channels.

In networks, like the EVN and the VLBA, that observe frequently in standard modes and in a stable configuration, the preparatory steps and correlation can be largely automated and pipelined. 
In other instances and especially when there are technical changes at stations between sessions, or there is no automated collection of metadata (station clock models, actual tunings, updated coordinates, etc), the step of preparing initial correlator \texttt{VEX} and configuration files can involve extensive manual editing.
For \ac{vlbi} arrays that do not observe regularly with a consistent setup, multiple iterations of re-correlating plus data inspection on a small subset of scans are usually required, until one arrives at a final correlator configuration. 


\subsection{Correlator Output}

After a correlation run (bottom of \Cref{fig:correlatorFlow}) the output data are typically converted from a correlator-native format into the widely adopted \texttt{FITS-IDI}\footnote{\texttt{FITS-IDI}: \url{https://fits.gsfc.nasa.gov/registry/fitsidi/AIPSMEM114.PDF}.}, \ac{casa} \ac{ms}\footnote{\ac{casa} \ac{ms}: \url{https://casa.nrao.edu/Memos/229.html}.}, \ac{hops}\texttt{/MkIV}\footnote{\ac{hops}\texttt{/MKIV}: \url{https://www.haystack.mit.edu/wp-content/uploads/2020/07/docs_hops_002_mk4_files.txt}.}, or \texttt{vgosDB NetCDF}\footnote{\texttt{vgosDB NetCDF}: \url{https://vievswiki.geo.tuwien.ac.at/VLBI-Analysis/Input-data}.} visibility data formats.
Metadata or single-dish measurements (top of \Cref{fig:correlatorFlow}) may further be attached to the visibility data-set as calibration tables. These may include system temperatures, antenna gain-elevation characteristics, weather data, flagging tables, and phase cal tones.

\ac{vlbi} arrays typically observe in a circular polarization basis (through the use of quarter-waveplates), as the calibration of data from linear feeds is coupled to the different feed angle rotations at different antennas in a complicated way.
If observations had some telescopes observing in linear rather than circular polarization, the visibilities will have a mixed polarization basis, which complicates the analysis of the data and is not supported by the \texttt{FITS-IDI} format. The \href{https://earth.gsfc.nasa.gov/geo/instruments/vlbi-global-observing-system-vgos}{VLBI Global Observing System} (VGOS), some EVN, and also (sub)mm-VLBI observations that include the phased Atacama Large Millimeter/submillimeter Array \citep[ALMA,][]{2009IEEEP..97.1463W,2018PASP..130a5002M} have to deal with this problem.
Before release to the PI, the mixed polarization visibilities are converted into a purely circular basis with the \texttt{PolConvert} \cite{2016A&A...587A.143M}\footnote{\url{https://github.com/marti-vidal-i/PolConvert}.} program. The conversion requires telescope D-term and cross-polarization phase and gain characterization (\Cref{sec:pol}). The ALMA Phasing Project quality-assurance (QA2) deliverables provide these parameters a~priori for ALMA \cite{2019PASP..131g5003G}. Alternatively, parameters can be derived from the visibilities themselves, via a \texttt{PolConvert} calibration pass on unpolarized sources.


\section{Signal stabilization}
\label{sec:signal}

As described above, the correlator delay models are not perfect, which leads to residual delays and rates being present in the visibility data. 
Additionally, corrections for atmospheric phase fluctuations and instrumental bandpasses, where telescopes exhibit phase and amplitude variations over their frequency response, are usually not applied at the correlator.
Post-correlation processing steps are therefore used to stabilize the source signal, ideally to remove all errors in the data down to the thermal noise.
This allows us to average the data significantly in time and frequency, with limits set only by the structure and variability of the observed source in the presence of thermal noise only.
In practice, the \ac{sn} of the data limits the degree to which data errors can be solved for.
Without the signal stabilization steps, baseline-based errors will be frozen into the data due to phase decoherence, delay errors, and imperfect bandpass responses for example.

Instrumental errors such as amplitude and phase bandpasses, phase and delay offsets between analog frequency bands, and polarization imperfections of the receivers (\Cref{sec:pol}), are stable or drift only slowly over the course of observations.
\ac{vlbi} scans on bright calibrators can thus be used to solve for these effects.
Some telescopes inject and log phase/pulse cal signals into their signal chain (\Cref{sec:corrparams}). These signals appear as weak periodic spectral lines (and should therefore not be used for spectral line observations) and can be used to estimate instrumental delay errors.

A common fringe-fitting method is to pick a reference station, transform the visibilities on baselines to that station into the delay-rate space with a fast Fourier transform (FFT), and to pick delay and rate estimates at the peak location in the two-dimensional FFT space \cite{alef1986}.
The delay, rate, and phase is set to zero for the reference station and the FFT peaks are used as starting guesses for the station-based delays and rates of all other antennas.
Additionally, the FFT \ac{sn}s (\Cref{tab:conventions}) can serve as a threshold to distinguish between detections and non-detections.
Finally, the data from all baselines on which the source is detected are used to refine the starting guesses and obtain final delay, rate, and phase solutions with a least-squares solver based on an assumed source model \cite{Schwab1983}.
The default assumption of a point source models usually does not lead to noticeable errors compared to the thermal noise \citep{2019ApJ...875L...3E}.
Exceptions are highly sensitive arrays with a low thermal noise or geodesy experiments, where source structures might prevent reaching the required accuracy on the delays.
If necessary, an iterative approach can be employed, where an image is first reconstructed (\Cref{sec:image}) using data calibrated with the assumption of a point source, that image is then used as source model to fringe-fit the data again, and the data with this improved calibration is then used to create a final image.
Alternatively, it is possible, although computationally very expensive, to solve for the combined fringes and source structure using a Bayesian approach \citep{2020Natarajan}.

Unlike instrumental errors, the residual post-correlation delays, rates, and atmospheric phase fluctuations are source/direction-dependent.
In cm-\ac{vlbi} experiments, atmospheric errors are moderate and calibration solutions can be transferred from nearby calibrators to the science target.
The closer the calibrator is to the science target, the better are the results; ideally both are within the same telescope primary beams. If the in-beam calibrator is weak, a ``bi-gradient phase referencing'' \citep{2006Doi} can be employed, where a further away very bright source is used to detect a closer weak calibrator via phase-referencing, from which phase-referencing is then used to detect the very weak science target.
Two col-linear aligned calibrators on opposite sides of a target source can be used for an improved correction of atmospheric errors \citep{2002Fomalont}. With three or more calibrators, a two-dimensional phase screen of the atmosphere can be corrected for via two-dimensional interpolation of antenna phases using a ``MultiView'' technique \citep{2017Rioja, 2022Hyland}.
At high observing frequencies, dominant path delay uncertainties are caused by the troposphere. 
Akin to geodesy \ac{vlbi}, ``geodetic blocks'' of $\sim$\,10 sources can be observed in short succession to accurately determine the zenith tropospheric delays at each telescope \citep{ReidHonma2014}.
Phase-referencing allows for astrometric measurements of the science target with respect to calibrator sources.
At mm wavelengths, the science target itself has to be fringe-fitted and must therefore be bright enough to be detected within atmospheric coherence times or incoherent averaging has to be used \citep{1995Rogers}.

With multi-frequency receivers, it is possible to transfer phase solutions across bands to extend the coherence time at the higher frequencies and enable the detection of weaker sources.
A simple frequency-phase-transfer (FPT) removes short-timescale tropospheric phase errors but does not allow for astrometric source registration.
This technique can be combined with a phase-referencing between multiple targets as ``source frequency phase referencing'' \citep[(SFPR),][]{2008Rioja, 2011Rioja, 2014Rioja, 2015Rioja}.
SFPR works at mm wavelengths, removes slowly varying residual ionospheric and instrumental effects not captured by the FPT, enables astrometry, and allows for longer on-source \ac{vlbi} scans as well as longer telescope slewing times to calibrator sources that are further away, since the fast tropospheric phase errors have been removed by the FPT.
\citet{2020Rioja} provide an outlook of high-precision astrometry measurements that will be enabled with future instruments and surveys.

\begin{figure}
\flushleft
\begin{tikzpicture}
  \def \yspc {-2cm}
  \def \yitemspc {-0.20cm}
  \def \yitemspcB {-0.40cm}
  \def \xpsc {3cm}
  \tikzstyle{block} = [
    draw=black,
    minimum width = 3.0cm, minimum height = 1.0cm,
    align = center, thick, rounded corners=.25cm
    ]
  \tikzstyle{blockitem} = [
    minimum width = 3.0cm, minimum height = 1.0cm,
    align = left, thick, rounded corners=0cm,
    text width=12cm, node distance=5cm
    ]
  \tikzstyle{blockoptionalitem} = [
    draw=black,dashed,
    minimum width = 3.0cm, minimum height = 1.0cm,
    align = center, thick, rounded corners=.25cm,
    text width=3.5cm
    ]
    \tikzstyle{arrow} = [thick, ->]

  \node[block, fill=orange!30] (P1) {Correlation};
  \node[blockitem, right of=P1, xshift=\xpsc] (P1a) {See \Cref{fig:correlatorFlow}.};

  \node[block, fill=green!10, below of=P1, yshift=-0.7cm] (P2) {Load visibilities \\ and metadata};
  \node[blockitem, right of=P2, xshift=\xpsc] (P2a) {Load everything in the \ac{ms} format, e.g., source models, ANTAB data, and weather information.};
  \draw[arrow] (P1) -- (P2);

  \node[block, fill=yellow!30, below of=P2, yshift=-0.7cm] (p1) {Diagnostics};
  \node[blockitem, right of=p1, xshift=\xpsc] {Inspect the raw data.};
  \draw[arrow] (P2) -- (p1);

  \node[block, fill=green!10, below of=p1, yshift=-0.7cm] (P3) {Flagging};
  \node[blockitem, right of=P3, xshift=\xpsc] (P3a) {Using a priori flags from the correlator and station log files.};
  \draw[arrow] (p1) -- (P3);

  \node[block, fill=blue!10, below of=P3, yshift=-0.7cm] (P4) {Autocorrelation-\\based calibration};
  \node[blockitem, right of=P4, xshift=\xpsc] {Correct sampler thresholds errors (\texttt{accor}) and perform scalar bandpass correction in the absence of strong \ac{rfi}.};
  \draw[arrow] (P3) -- (P4);
  
  \node[block, fill=blue!10, below of=P4, yshift=-0.7cm] (P5) {Flux density \\ calibration};
  \node[blockitem, right of=P5, xshift=\xpsc] {Using ANTAB data (\Cref{sec:flux}).};
  \draw[arrow] (P4) -- (P5);
 
  \node[block, fill=red!20, below of=P5, yshift=-0.7cm] (P6) {Atmospheric \\ pre-calibration};
  \node[blockitem, right of=P6, xshift=\xpsc] {\textit{Low $\nu$}\,($ \lesssim 20$\,GHz): \ac{tec} ionospheric calibration.\\
                                               \textit{High $\nu$}\,($ \gtrsim 20$\,GHz): Segmented fitting of calibrator phases and rates to correct troposheric turbulence on short timescales.};
  \draw[arrow] (P5) -- (P6);

  \node[block, fill=red!20, below of=P6, yshift=-0.7cm] (P7) {Instrumental \\ calibration};
  \node[blockitem, right of=P7, xshift=\xpsc] {Correct phase and delay offsets between spectral windows, calibrate complex bandpass response, and solve reference antenna R-L delay. If the measurements are available, set overall R-L phase and correct leakage.};
  \draw[arrow] (P6) -- (P7);

  \node[block, fill=red!20, below of=P7, yshift=-0.7cm] (P8) {Fringe-fit};
  \node[blockitem, right of=P8, xshift=\xpsc] {Fringe-fit over scan durations. \textit{Low $\nu$}: Phase-referencing if needed.\\
  \textit{High $\nu$}: Additional segmented fringe-fitting in a second stage.};
  \draw[arrow] (P7) -- (P8);
  
  \node[block, fill=green!10, below of=P8, yshift=-0.7cm] (P9) {Calibrate\\visibilities};
  \node[blockitem, right of=P9, xshift=\xpsc] {Apply all incremental calibration tables.};
  \draw[arrow] (P8) -- (P9);
  
  \node[block, fill=yellow!30, below of=P9, yshift=-0.7cm] (P10) {Diagnostics};
  \node[blockitem, right of=P10, xshift=\xpsc] {Inspect the calibration solutions and the calibrated data.};
  \draw[arrow] (P9) -- (P10);

  \node[block, fill=green!10, below of=P10, yshift=-0.7cm] (P11) {Data output};
  \node[blockitem, right of=P11, xshift=\xpsc] {Average visibilities in appropriate time and frequency bins. Export the calibrated data in the \ac{ms} and/or UVFITS format.};
  \draw[arrow] (P10) -- (P11);

  \node[block, fill=orange!30, below of=P11, yshift=-0.7cm] (P12) {Science};
  \node[blockitem, right of=P12, xshift=\xpsc] {Imaging, model-fitting, data analysis, publication.};
  \draw[arrow] (P11) -- (P12);

\end{tikzpicture}
\caption{Flowchart of typical continuum \ac{vlbi} data reduction steps shown in the order as implemented in the \texttt{rPICARD} pipeline (\Cref{sec:rpicard}). Feed rotation angle effects are calibrated on-the-fly here.}\label{fig:calibrationrFlow}
\end{figure}
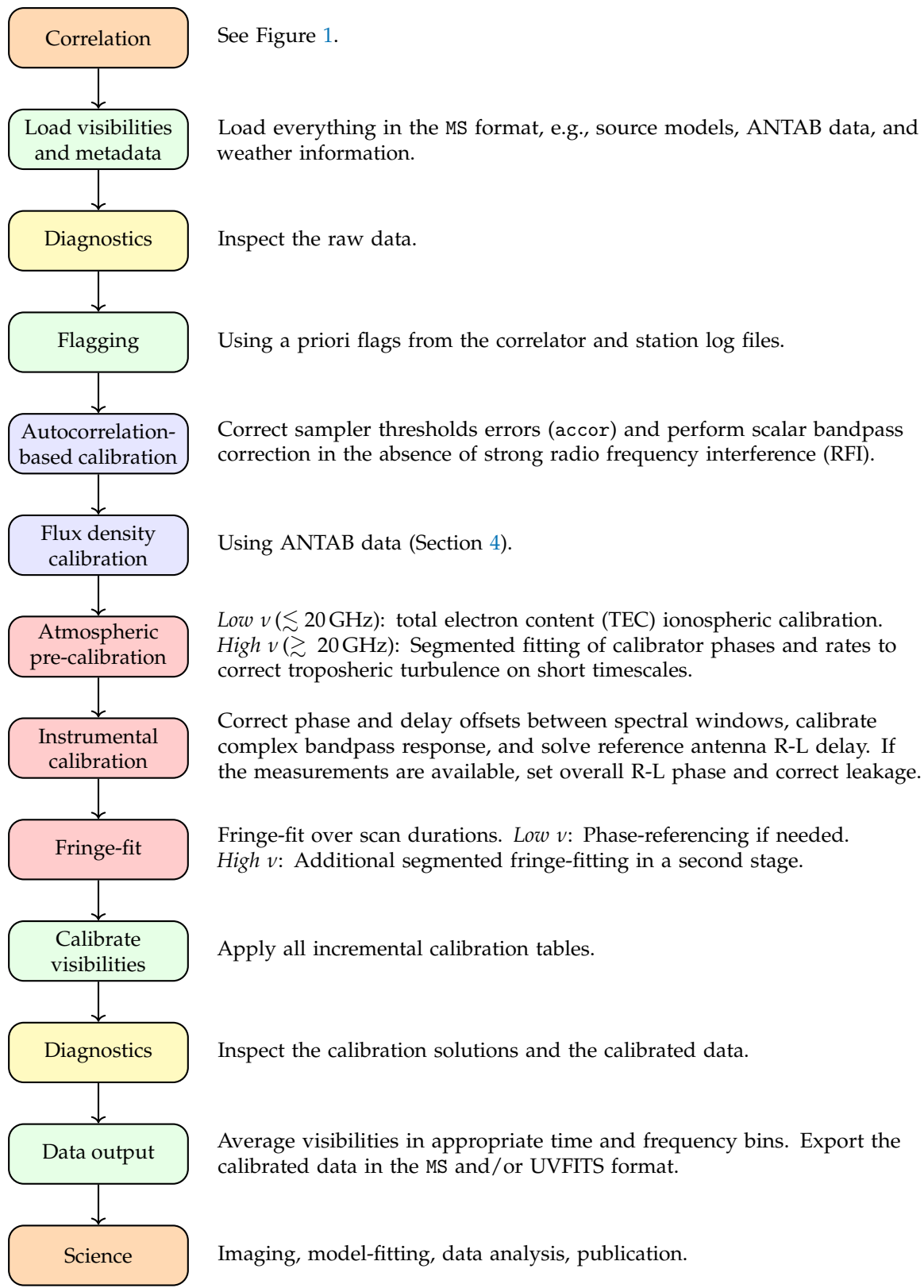


The complex post-correlation data in its multi-dimensional polarization (correlation products), frequency, time, and baseline space has a considerable size and is challenging to handle.
Particular difficulties can arise in heterogeneous \ac{vlbi} arrays, where different stations may require different treatments, results may be dominated by the measurements (and potential systematics) of the most sensitive telescope(s), and varying station sensitivities may make a uniform data reduction difficult. For example, calibration solutions with a high time-and/or frequency resolution can be obtained for the sensitive telescopes, while large solution windows might be required to accumulate enough \ac{sn} for the less sensitive telescopes, even when all baselines are used. For these considerations, array geometry also plays an important role; high \ac{sn} data from the large scale emission of resolved sources can be obtained for smaller dishes if they have short baselines to more sensitive telescopes.

Errors in the data, which can typically be traced back to single antennas, are often not trivial to find; a variety of diagnostic tools and visualizations that make different cuts through the data are needed. Some -- but typically not all -- errors can be identified at the signal stabilization step. \Cref{sec:flagging} provides an overview of how bad data can be identified and dealt with at the different data reduction stages.
Calibration errors are easily baked into the data and can have substantial effects on the final scientific results.
A typical order of \ac{vlbi} calibration operations are shown in \Cref{fig:calibrationrFlow}.
Comprehensive and well-tested software suites are generally used for the primary calibration steps.
The most common software packages are described below. Note that tools for the flux density calibration (\Cref{sec:flux}), imaging (\Cref{sec:image}), as well as advanced calibration methods for polarimetry (\Cref{sec:pol}), spectral lines (\Cref{sec:line}), and wide fields of view (\Cref{sec:wide}) are implemented in several of these packages.

\subsection{AIPS}

The \ac{aips} \citep{2003Greisen}\footnote{\url{http://www.aips.nrao.edu}.} is currently the most-widely used tool for the post-correlation calibration of \ac{vlbi} data.
The common visibility input data format that is produced by modern \ac{vlbi} correlators is \texttt{FITS-IDI}.
The calibrated output data are saved as \texttt{UVFITS}\footnote{The \texttt{UVFITS} format is following \ac{aips} conventions (\url{ftp://ftp.aoc.nrao.edu/pub/software/aips/TEXT/PUBL/AIPSMEM117.PS}), which can sometimes change. The format is therefore not well defined and different software packages use slightly different conventions.} file to disk.

With \ac{aips}, ionospheric corrections based on a priori maps of the global \ac{tec} can be applied to the data for an improved astrometry.
Furthermore, the applied fringe corrections from the correlator can be improved with a posteriori determined \acp{eop}.
Finally, pulse cal signals can be used to calibrate instrumental effects.

Several hundred tasks are available within \ac{aips}, covering all standard data calibration and analysis steps as well as advanced methods, such as computing the statistics on Allan Variances (\texttt{ALVPR}~task), modeling of a gravitational lenses (\texttt{GLENS}), and optimizing configurations of telescope arrays (\texttt{CONFI}), to name a few examples.

The \ac{aips} cookbook\footnote{\url{http://www.aips.nrao.edu/cook.html}.} can serve as a detailed guide to typical VLBI data reduction pathways, where the needed \ac{aips} tasks and their parameters are described for each step. Additionally, several generically useful \ac{vlbi} tips and tricks are given.

\subsubsection{ParselTongue}

\href{https://www.jive.eu/jivewiki/doku.php?id=parseltongue:parseltongue}{\texttt{ParselTongue}} \citep{2006Kettenis}\footnote{\url{https://www.jive.eu/jivewiki/doku.php?id=parseltongue:parseltongue}.} is a \texttt{Python} interface for \ac{aips}. It allows the user to run \ac{aips} tasks, access calibration and visibility data, and to script their data processing. A \texttt{ParselTongue} pipeline is currently used for EVN pipelining operations.

\subsection{CASA}

The \ac{casa} software \citep[][Emonts et al. in prep.]{2007ASPC..376..127M}\footnote{\url{https://casa.nrao.edu}.} and its underlying suite of \texttt{C++} ``\texttt{casacore}'' libraries are widely used for the processing of connected-element-interferometer radio data (e.g., from ALMA and the VLA) and also single-dish radio observations. \citep[e.g.,][]{1980ApJS...44..151T,2009IEEEP..97.1463W,2009IEEEP..97.1482D,2013A&A...556A...2V}.
Initiated by the ``BlackHoleCam'' project \citep{2017IJMPD..2630001G}, \ac{casa} has been recently upgraded with \ac{vlbi} capabilities, which are now supported and further developed by the \ac{jive} and \ac{nrao} \citep{2022Bemmel}.
The input visibility can be in the native, locally stored \ac{ms} format or \texttt{FITS-IDI} files, which will be loaded and converted to a \ac{ms}.
The final calibrated visibilities are stored together with the raw data in the \ac{ms} and can also be exported as \texttt{UVFITS} files.

The \ac{casa} calibration framework is based on the Hamaker-Bregman-Sault measurement equation, a powerful mathematical framework for interferometric calibration problems, including direction-dependent effects \citep{1996A&AS..117..137H, oms2011, oms2011a, oms2011b, oms2011c}.
Unlike \ac{aips}, \ac{casa} stores the visibility data (in \texttt{MS} format) and calibration tables (in a \texttt{MS}-like format) separately on disk in the working directories.

The fringe-fitter and all standard \ac{casa} tasks can make use of a source model to improve calibration accuracy. The source model information is stored in the \ac{ms} when the imaging is done within \ac{casa} (\Cref{sec:image}) or when an external \texttt{FITS} image is loaded.
Ionospheric \ac{tec} corrections can also be done in \ac{casa}.
Additionally, the fringe-fitter has the capability of solving for dispersive delays from the ionosphere.

Development plans for the short-term future include the handling of pulse cal data, an improved handling of irregularly spaced frequency channels when performing multi-band fringe-fitting, \ac{eop}s correction and a general accountability of the correlator model \citep{2022Bemmel}.

The next-generation \ac{casa} (ng\ac{casa}) project\footnote{\texttt{ng}\ac{casa}: \url{https://pypi.org/project/ngcasa}.} will improve the hardware scalability significantly, particularly the speed with which large data volumes from future instruments can be accessed and processed, by using off-the-shelf \texttt{Python} libraries like \texttt{Dask} \citep{dask} and \texttt{xarray} \citep{xarray}.

\subsubsection{rPICARD}
\label{sec:rpicard}

\ac{casa} has a convenient toolkit for extensive visbility and calibration data access, which allows users to easily automate their data processing.
This led to the development of a generic \ac{vlbi} calibration pipeline, the \ac{rpicard}, which is based on \ac{casa} \citep{2018evn..confE..80J,rpicard,2019ApJ...875L...3E}.
\ac{rpicard} employs flexible calibration strategies that work for any array and also synthetic \ac{vlbi} data \citep{2017Blecher,2022Natarajan}.

\begin{figure}
\centering
\includegraphics[width=0.8\columnwidth]{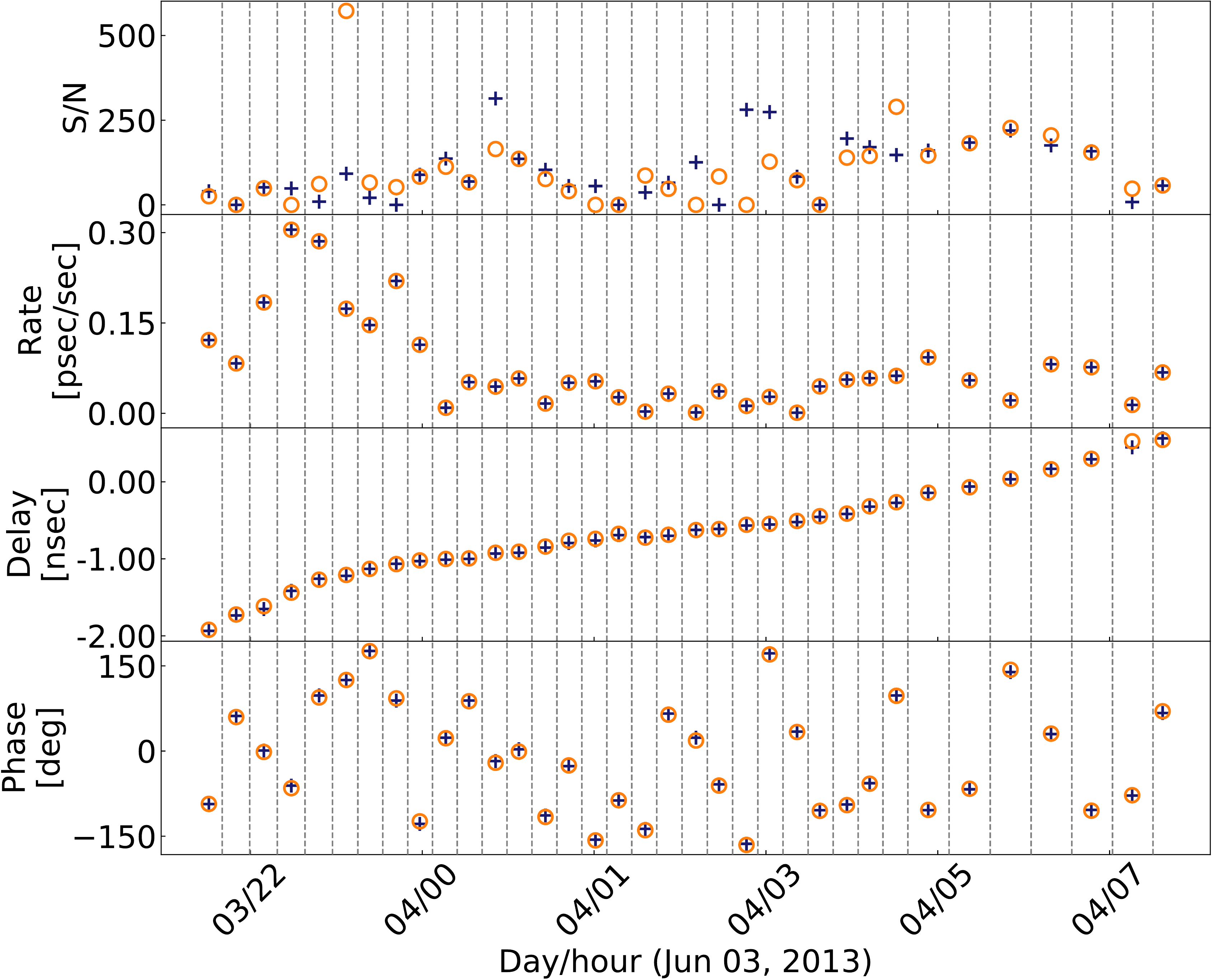}
\includegraphics[width=0.8\columnwidth]{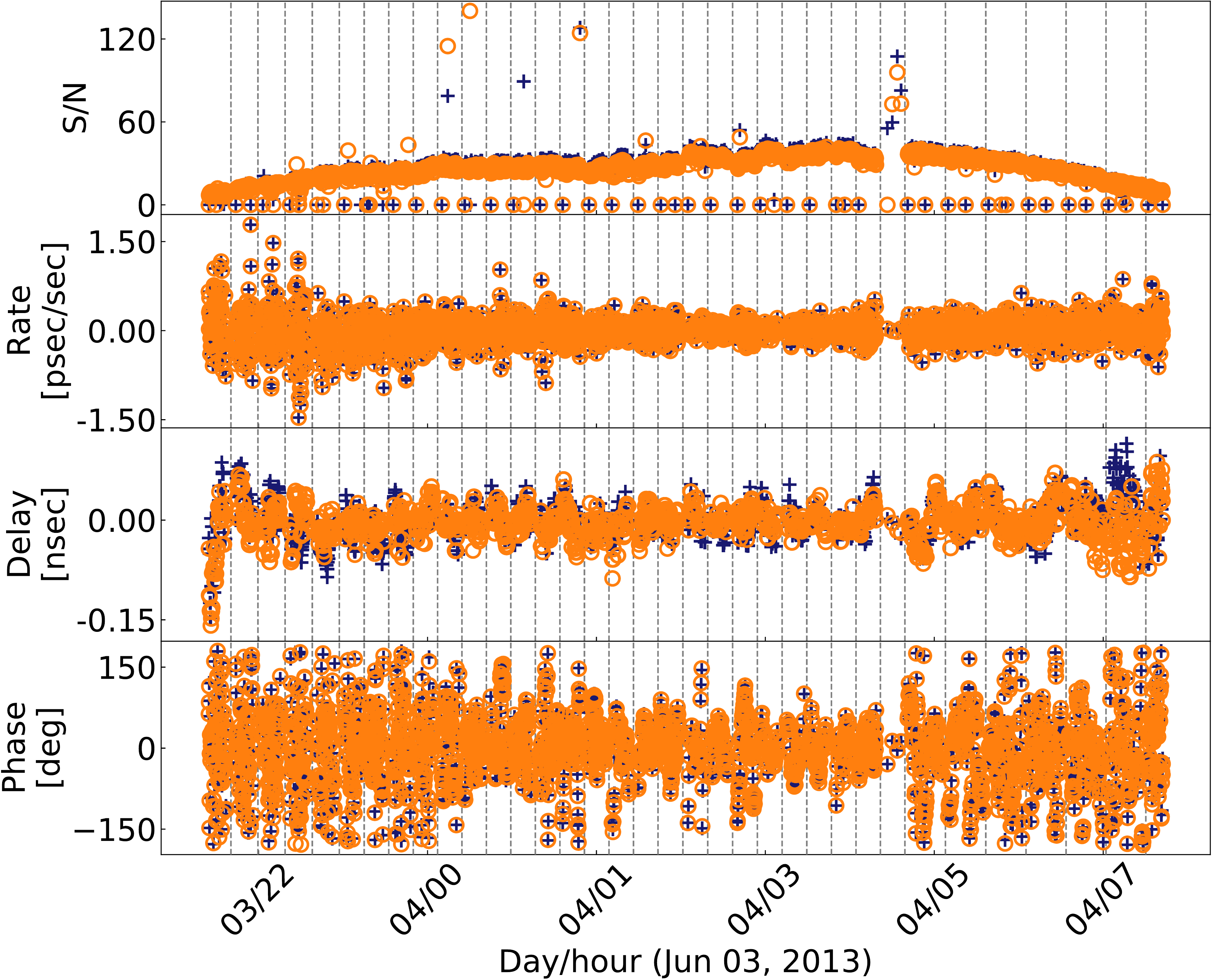}
\caption{Example fringe solution diagnostics from \ac{rpicard} of M87 observed at 43\,GHz with the VLBA (project code BW0106). Dashed lines show scan boundaries. Crosses and circles correspond to RCP and LCP, respectively. \textit{Top}: Gross per-scan solutions. \textit{Bottom}: Residual atmospheric corrections on adaptive segmentation timescales after the solutions from the top panel have been applied to the data.}
\label{fig:calib}
\end{figure}

\ac{rpicard} uses MPI for scalability and a custom robust scalar bandpass calibration method.
The ionospheric dispersive delay and Faraday rotation correction is done with standard \ac{casa} tasks using \ac{tec} files from NASA's Crustal Dynamics Data Information System.
Fringe-fitting is done with the \ac{casa} Schwab-Cotton algorithm \cite{Schwab1983} and a simple single-source phase referencing method is implemented in \ac{rpicard}. Instrumental phase, delay, and bandpass effects are corrected by utilizing all scans on calibrator sources. For corrections of atmospheric phase turbulence at high observing frequencies, the fringe-fitting is done on a segmentation time, which is optimized based on the expected atmospheric coherence time at the observing frequency and the \ac{sn} of the fringe detections. \Cref{fig:calib} shows this two-step fringe-fitting approach, where fringe detections are first obtained over \ac{vlbi} scan durations and residual atmospherically-induced errors are subsequently calibrated out on short segmentation times.

Next to the aforementioned methods, which are implemented through standard \ac{casa} methods, \ac{rpicard} also has a few custom capabilities: 1) Exhaustive baseline searches to an arbitrary order are performed for non-detections to a primary reference antenna \citep[Appendix D in ][]{rpicard}. 2) The same rate solutions are applied to all polarization channels and a simple nearest-neighbor algorithm is used to re-reference all fringe solutions to a single antenna for the stability of polarized source signals.
3) Robust solutions for atmospheric opacities are derived with estimations of the atmospheric temperature from the \texttt{ATM} software \citep{2001ITAP...49.1683P}.

\ac{rpicard} can be obtained from a Bitbucket repository\footnote{\url{https://bitbucket.org/M_Janssen/picard}.} and for every version of the pipeline, \texttt{Docker} and \texttt{Singularity} containers are available (details are given in the \texttt{README} file of the pipeline repository).

\subsubsection{VPIPE}

The \texttt{VLBI PIPEline (VPIPE)} pipeline is an end-to-end \ac{casa}-based VLBI pipeline designed for mid-frequency ($\sim\mathrm{GHz}$) observations\footnote{\href{https://github.com/jradcliffe5/VLBI_pipeline}{https://github.com/jradcliffe5/VLBI\_pipeline.}}. It currently supports the EVN, VLBA and LBA networks. The pipeline employs similar steps to the \ac{rpicard} pipeline (\S\,\ref{sec:rpicard}) including TEC corrections, phase referencing via fringe fitting, and self-calibration. 

Originally designed for the high data volumes associated with wide-field VLBI data (typically 10--50\,TB), the pipeline utilises \texttt{MPI-CASA} and HPC parallelism provided by cluster scheduling software, such as \texttt{SLURM} and \texttt{PBS Pro}, for enhanced scalability. This allows the pipeline to process data quickly and efficiently, and permit the processing of ``multiple phase centre observing'' correlated data within short timescales (see \S\,\ref{sec:wide}) \citep[][]{2011PASP..123..275D}. \texttt{VPIPE} is designed to be highly modular and supports containerization technologies, such as \texttt{Singularity}. This enables the user to easily add extra processing/analysis steps that uses custom scripts or software packages.

The pipeline incorporates advanced direction-dependent calibration steps needed for wide-field VLBI data. These include multi-source self-calibration (MSSC; see \S\,\ref{sec:wide}), and primary beam correction schemes. This can be conducted in the \uv{} or image plane. Correcting for these allows the pipeline to reduce noise levels, improve the dynamic range, increase the effective field-of-view, and ensure consistent flux density scales across the primary beam.

\subsubsection{The \textit{e}-MERLIN CASA Pipeline}

The \texttt{\textit{e}-MERLIN \ac{casa} Pipeline (eMCP)} is a \texttt{Python} pipeline that is designed for use with \textit{e}-MERLIN data\footnote{\href{https://github.com/e-merlin/eMERLIN_CASA_pipeline}{https://github.com/e-merlin/eMERLIN\_CASA\_pipeline.}}. It currently supports continuum and spectral line observations. The pipeline uses \ac{casa}-based VLBI tools, such as baseline and global fringe-fitting routines, to perform phase-referencing and self-calibration with parameters that are tailored for the \textit{e}-MERLIN array and the observing frequency. The pipeline adopts other software tools such as \texttt{AOFlagger} \citep{2012A&A...539A..95O}\footnote{\url{https://gitlab.com/aroffringa/aoflagger}.} and \texttt{WSClean} \citep{2014MNRAS.444..606O}\footnote{\url{https://gitlab.com/aroffringa/wsclean}.} to address the challenging RFI environment and large image sizes often required. 

Since 2019, eMCP has been incorporated as the official pipeline for the processing and delivery of \textit{e}-MERLIN data to the users. As a result, it contains a suite of advanced quality assessment tools, displayed through a web-interface, that permit the user easily assess the quality of the calibration and images of the science targets.

\subsubsection{JIVE EVN continuum Jupyter notebooks}

EVN users are provided with a \texttt{ParselTongue}-based pipeline-calibrated version of their data by \ac{jive}.
Additionally, a \ac{casa}-based pipeline for EVN continuum is under development at \ac{jive}, which will run in a Jupyter notebook \citep{2020Keimpema, 2022Bemmel}\footnote{\url{https://code.jive.eu/bemmel/EVN_CASA_pipeline}. See also \url{https://www.evlbi.org/evn-data-reduction-guide}.}
With this notebook, users will be able to start with a default set of parameters and steps used to calibrate EVN data, which can then be adjusted and refined to produce science-ready data.

\subsection{EHT-HOPS}

\ac{hops}\footnote{\url{https://www.haystack.mit.edu/haystack-observatory-postprocessing-system-hops}.} is a collection of basic \ac{vlbi} functions like fringe-fitting, which are sufficient for the current needs of geodetic analyses \citep{2004RaSc...39.1007W}.
However, \ac{hops} is not as flexible as \ac{aips} and \ac{casa} and lacks a few capabilities, which prevents a full scientific data analysis.
The missing features will be implemented in an updated next-generation \ac{hops} (\texttt{ng}\ac{hops}) software package.

In the meantime, an ``\texttt{EHT-HOPS}'' \citep{2019ApJ...882...23B, 2019ApJ...875L...3E} pipeline has been designed, which augments \ac{hops} with custom \texttt{Python} scripts\footnote{\url{https://github.com/sao-eht/eat}.} to enable the full reduction of data from the EHT and \ac{gmva} when ALMA is participating in the observations \citep{2019ApJ...871...30I, 2019ApJ...875L...3E}.
A least-squares solver is used to find station-based delay and rate solutions from the baseline-based estimates that \ac{hops} can solve for \cite{alef1986}.
High \ac{sn} detections on bright calibrators are used to solve for bandpass effects.
A piece-wise polynomial phase model is fitted directly to the visibilities to correct for atmospheric phase turbulence. To avoid over-fitting, phase corrections for each spectral window are obtained from a fit to all other spectral windows in the data.
Relative complex gains between polarization channels are fitted to facilitate polarimetric analyses.

\texttt{EHT-HOPS} and \ac{rpicard} are used to produce science-ready data from the EHT \citep{2019ApJ...875L...3E, Kim_2020, 2021Janssen, eht-SgrAii}.


\subsection{PIMA}

\href{http://astrogeo.org/pima}{\texttt{PIMA}} \citep{2011AJ....142...35P}\footnote{\url{http://astrogeo.org/pima}.} is a specialized data reduction program, which can be used to accurately distinguish fringe detections from non-detections.
\texttt{PIMA} is primarily employed for space VLBI experiments, because its fringe-fitter can solve for an acceleration (second-order rate) term.
As space VLBI is beyond the scope of this work, we will not consider \texttt{PIMA} further.


\section{Flux density calibration}
\label{sec:flux}

After the signal stabilization, the correlation coefficients in units of thermal noise are normalized to unity autocorrelations and scaled corresponding to idealized analog correlation amplitudes.
Ignoring baseline-dependent amplitude losses (such as de-correlation from data averaging), visibilities $\mathcal{V}_{ij}$ can be obtained in flux density units of \ac{jy} by calibrating the complex correlation coefficients $r_{ij}$ of the baseline connecting stations $i$ and $j$ with the a priori estimated \ac{sefd} sensitivities of our antennas,
\begin{equation}
\mathcal{V}_{ij} = \frac{1}{\eta_Q} \sqrt{\mathrm{SEFD}_i \mathrm{SEFD}_j e^{\tau_i + \tau_j}} r_{ij} \; .
\label{eq:apcal}
\end{equation}
Here, $\eta_Q$ is a correction factor for the digital quantization efficiency \citep{TMS}; for 2-bit sampling, we have $\eta_Q \approx 0.88$.
To correct for atmospheric attenuation of ground-based millimeter and submillimeter observations and thereby measure $\mathcal{V}_{ij}$ ``above the atmosphere'', $e^{\tau}$ correction factors are applied. Here, $\tau_i$ is the mean atmospheric opacity along the line of sight signal path of antenna $i$.

\begin{figure}
\centering
\includegraphics[width=0.95\columnwidth]{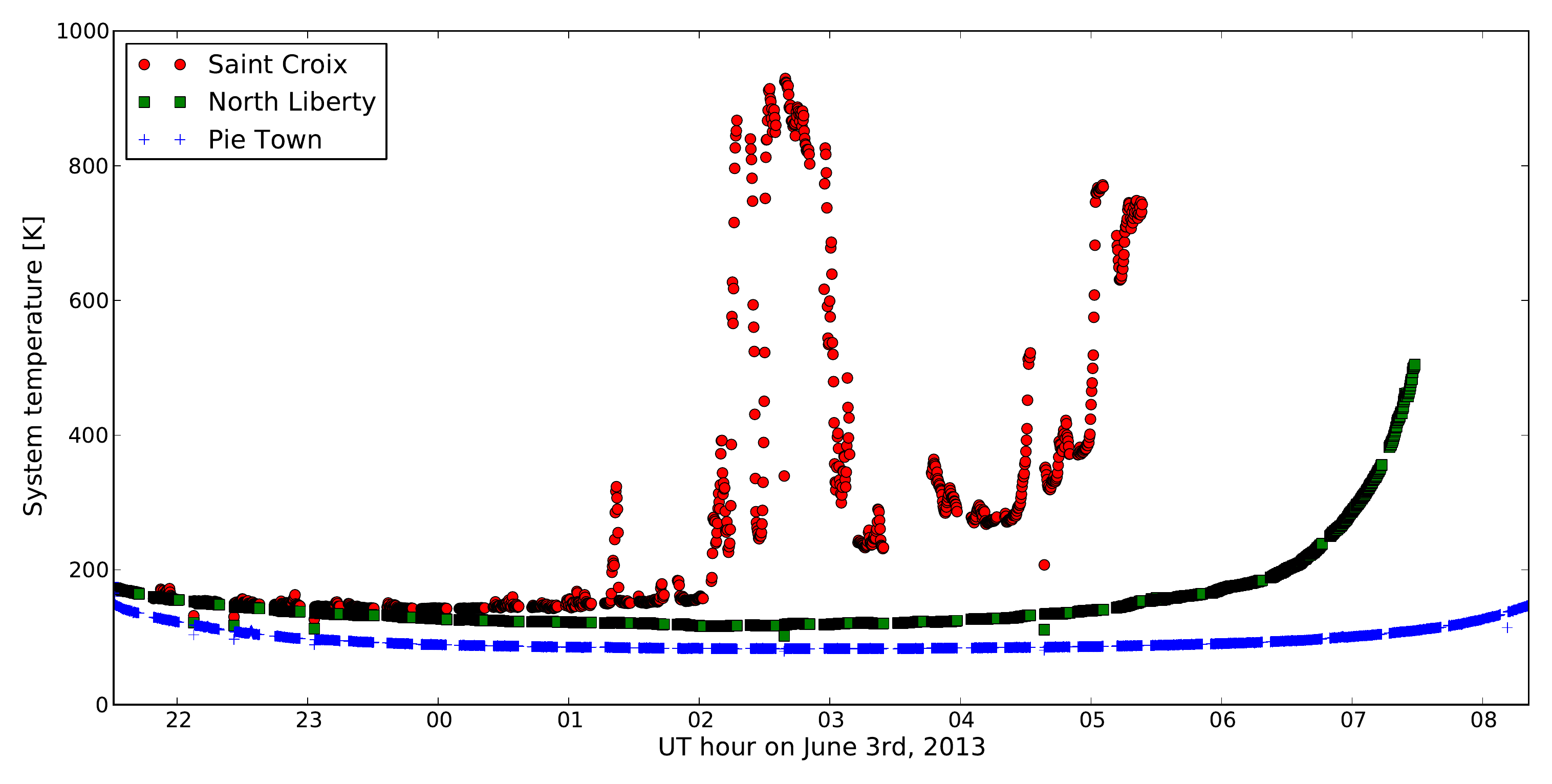}
\caption{Example of effective (opacity-corrected) system temperatures measured by three VLBA antennas from observations of M87 at 43\,GHz (project code BW0106). Pie Town system temperatures show slight increases at the beginning and end of the observing track, where the source is at a low elevation. The source sets at the end of the track for the North Liberty station, which causes an increases in system temperature. Bad weather is responsible for the highly variable system temperature measurements at Saint Croix.}
\label{fig:tsys}
\end{figure}

The \ac{sefd}s are based on measured system temperatures $T_\mathrm{sys}$, the telescope gain $\mathcal{G}$ as a function of elevation, and the phasing efficiency $\eta_\mathrm{ph}$ for phased arrays: $\mathrm{SEFD} \propto \eta_\mathrm{ph}^{-1} \mathcal{G}^{-1} T_\mathrm{sys}$.
Hot-load calibration scans measure the effective system temperature $T_\mathrm{sys} e^\tau$ directly and thereby immediately satisfy \autoref{eq:apcal} \citep{1976ApJS...30..247U}.
For noise-diode-based $T_\mathrm{sys}$ measurements, $\tau$ must be estimated after the fact.
A common approach is to first approximate the system temperature as $T_\mathrm{sys} \simeq T_\mathrm{rx} + \left(1-e^{-\tau}\right)T_\mathrm{atm}$, where $T_\mathrm{rx}$ and $T_\mathrm{atm}$ are the two dominant noise terms, corresponding to receiver and atmospheric temperatures, respectively.
Then, $T_\mathrm{rx}$ is estimated by extrapolating $T_\mathrm{sys}$ to zero airmass and an atmospheric model is used to estimate $T_\mathrm{atm}$ from weather parameters measured at the telescope. The $T_\mathrm{sys}$ measurements themselves can subsequently be used to estimate $\tau$ \citep[Section 4.3 in][]{rpicard}.
Examples of opacity-corrected system temperatures are shown in \Cref{fig:tsys}.

The calibration metadata, gathered as a priori gain and system temperature information from all telescopes, are commonly stored in simple ASCII text files following the \texttt{ANTAB} format\footnote{\texttt{ANTAB}: \url{http://www.aips.nrao.edu/cgi-bin/ZXHLP2.PL?ANTAB}.}. Note that invalid entries are often marked with a value of $999$ or $-999$.
Auxiliary weather information that might be needed to estimate $T_\mathrm{atm}$ for an opacity correction is usually attached to the visibility data format. Otherwise, it can be loaded from an ASCII text file (by \ac{aips} and \ac{rpicard} as described in \cite{rpicard}).

\subsection{Software implementations}

\ac{aips}, \ac{casa}/\ac{rpicard}, and \texttt{EHT-HOPS} post-processing scripts can be used to extract \ac{sefd}s from \texttt{ANTAB} tables and subsequently perform the a priori flux density calibration.
\ac{aips} reads external calibration information with the \texttt{ANTAB} task and creates \ac{sefd} calibration tables with \texttt{APCAL}, optionally with a simple fit for additional atmospheric opacity corrections.\footnote{\url{https://library.nrao.edu/public/memos/vlba/sci/VLBAS_01.pdf}.} Here, the measured ambient temperature $T_\mathrm{amb}$ is used to estimate the sky temperature via \mbox{$T_\mathrm{atm} = 1.12 T_\mathrm{amb} - 50\,\mathrm{K}$}.
\ac{casa} and \ac{rpicard} use custom \texttt{Python} modules\footnote{\url{https://github.com/jive-vlbi/casa-vlbi}.} to load the \texttt{ANTAB} data if it is not already attached to the \texttt{FITS-IDI} files. The \ac{casa} task \texttt{gencal} is used to create \ac{sefd} calibration tables (separately for telescope gains and system temperatures).
With a custom module, \ac{rpicard} can perform a robust atmospheric opacity correction for a user-specified selection of antennas. Here, the \texttt{ATM} code \citep{2001ITAP...49.1683P} is used to find $T_\mathrm{atm}$.
The \href{https://github.com/sao-eht/eat}{EHT Analysis Toolkit} provides post-processing scripts for the flux density calibration of \texttt{EHT-HOPS} data.

\subsection{Advanced methods based on array redundancy and a priori source assumptions}
\label{sec:special-calib}

\begin{itemize}
\item A \textit{network calibration} \citep{2019ApJ...882...23B, 2019ApJ...875L...3E} can be employed if the unresolved flux density on large scales seen by short baselines in the \ac{vlbi} array is known. The method only works for telescope sites that are close enough (almost co-located or in ``walking distance'') to effectively form a zero-baseline interferometer. The flux density measurement used to calibrate the gains can be obtained from (quasi-)simultaneous single-dish or connected-element-interferometry observations and allows for an absolute amplitude calibration. A least-squares approach with all baselines to pairs of redundant sites are used to robustly constrain the gains of the two neighboring antennas. A convenient implementation of the network calibration method can found in the \href{https://github.com/achael/eht-imaging}{\texttt{eht-imaging}} software (\Cref{sec:ehtim}).
\item \textit{Cross-track calibration} \citep{1984ARA&A..22...97P} is based on the fact that redundant baselines anywhere in the {\uv}-space should measure the same source properties modulo intrinsic source variability. 
A threshold can be set for how close two baselines should be to be considered identical, given how ``quickly'' the source structure varies in the Fourier space. Baselines might cross within a small region or stay closely parallel for long \uv-tracks. Network calibration is a special case of the cross-track calibration, where a known total flux density can be used for an absolute gain calibration. Generally, we have no a priori knowledge about the resolved source structure at larger \uv-spacings. Here, the least-squares solver can be used for all redundant baselines to tighten the gains of the involved stations by constraining the amplitude ratio of crossing-track baselines to unity. In absence of accurate and independent flux density information, the product of the solved gains are enforced to be unity to only solve for relative gains without adjusting the total flux density. Within \ac{aips}, the \texttt{UVCRS} task can be used as a diagnostic tool to identify anomalous gains of stations that are involved in crossing \uv-tracks. The only real cross-track calibration method implementation, that the authors are aware of, is the \href{https://sites.astro.caltech.edu/~tjp/citvlb/vlbhelp/uvcross.mem}{\texttt{UVCROSS Caltech VLBI Analysis Program}} \citep{1991BAAS...23..991P}.\footnote{The Caltech VLBI Analysis Programs (\url{https://sites.astro.caltech.edu/~tjp/citvlb}) have been discontinued and are therefore not considered further in this work.}
\item \textit{Second-moment source size calibration} \citep{2019A&A...629A..32I} can be employed under the assumption that the short baselines in the array sample a simple large-scale source structure such as a Gaussian. From the assumed large-scale image, model amplitudes can be computed and gains from the stations connected by short baselines can be obtained by self-calibration (\Cref{sec:image}). For a single baseline, a common application is to keep the gains fixed for the station which has the more accurate \ac{sefd}-based a priori flux density calibration. This method can be employed by all imaging software packages' self-calibration routines. A convenient implementation can be found in \texttt{eht-imaging}.
\end{itemize}


\section{Imaging and geometric model-fitting}
\label{sec:image}

\begin{figure}[t]
    \centering
    \begin{subfigure}[b]{0.5\textwidth}
	  \includegraphics[height=10.6cm]{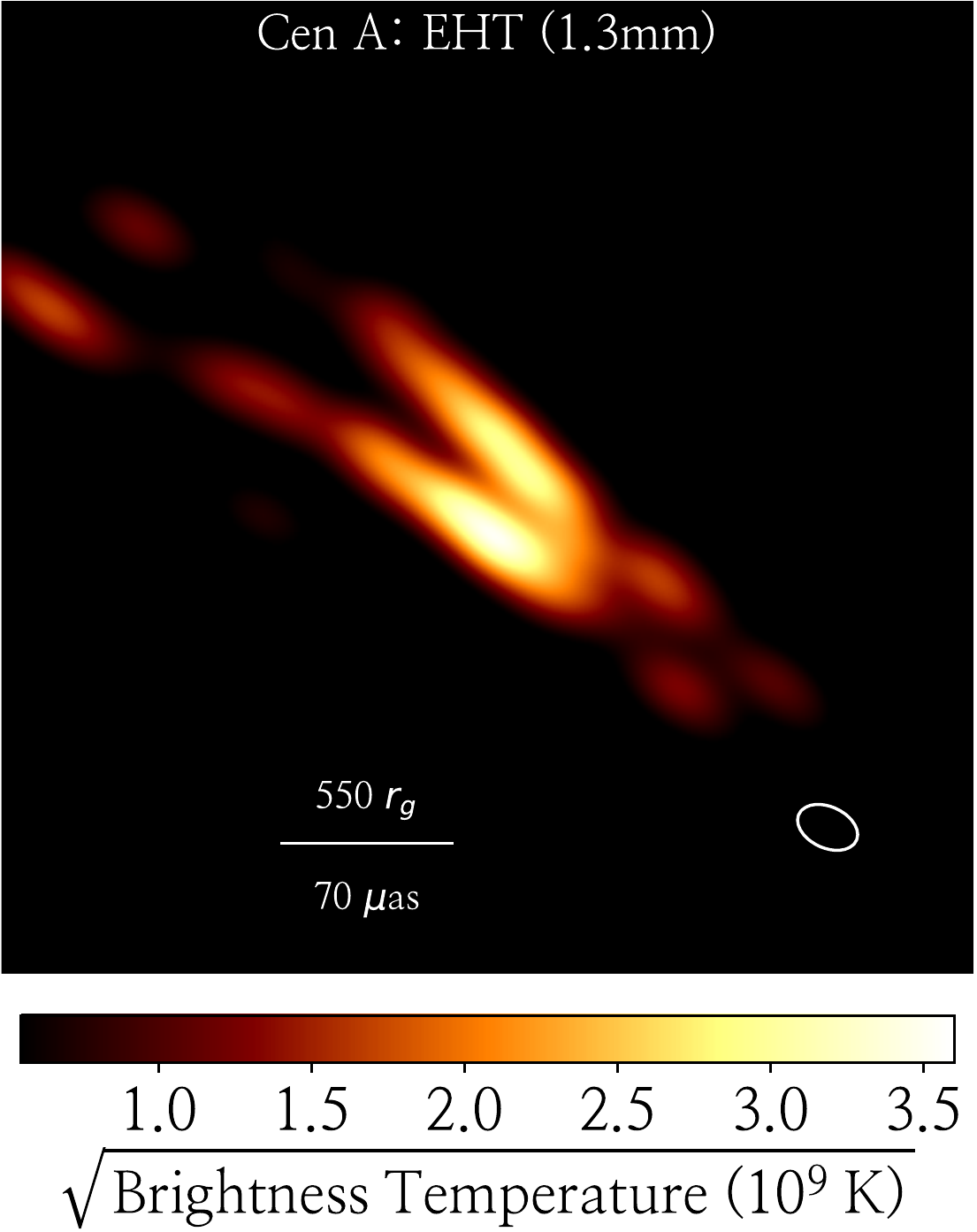}
    \end{subfigure}
    \hfill
    \begin{minipage}[b]{0.45\textwidth}
      \begin{subfigure}[b]{\linewidth}
	    \includegraphics[height=5.3cm]{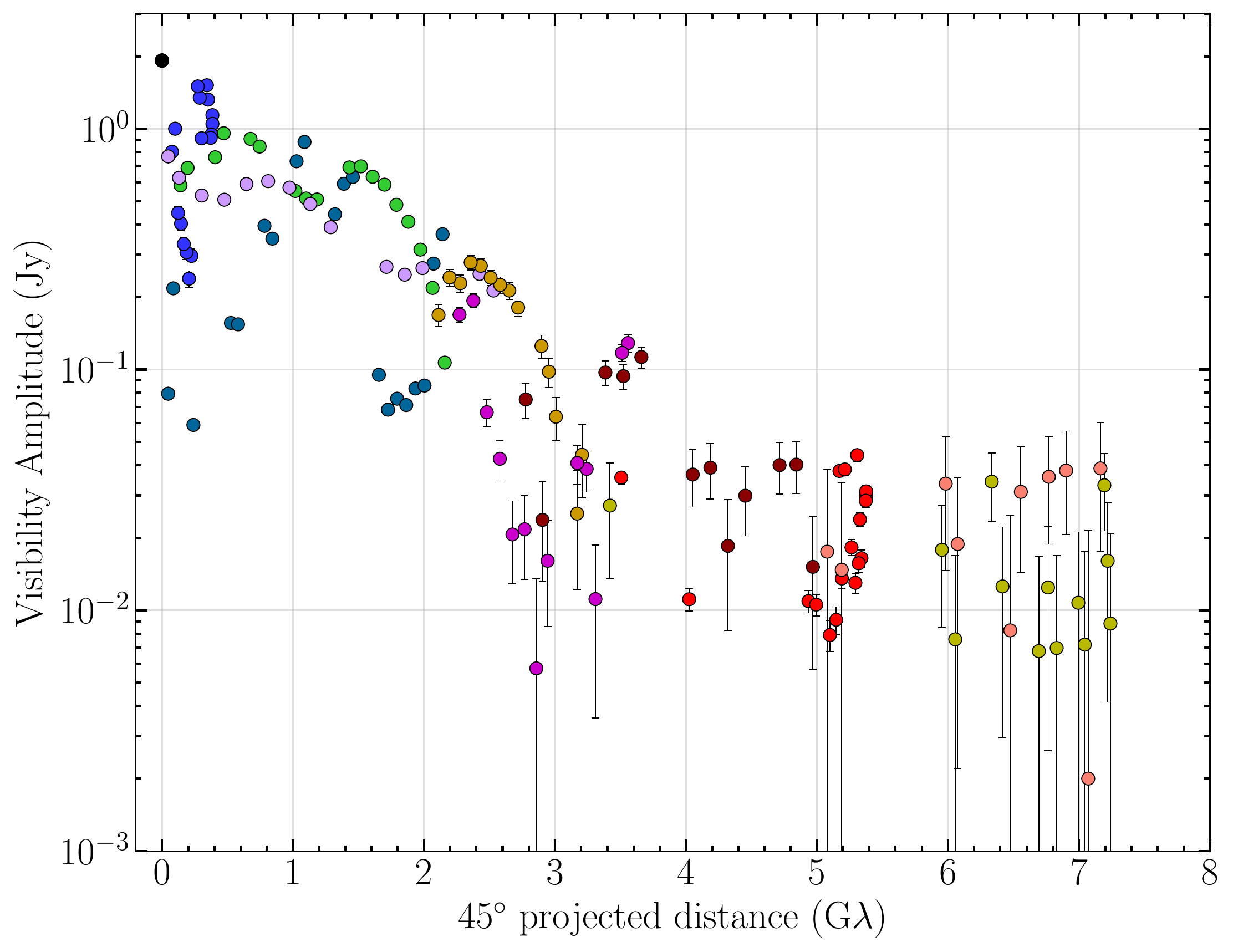}
      \end{subfigure}\\
      \begin{subfigure}[b]{\linewidth}
	    \includegraphics[height=5.3cm]{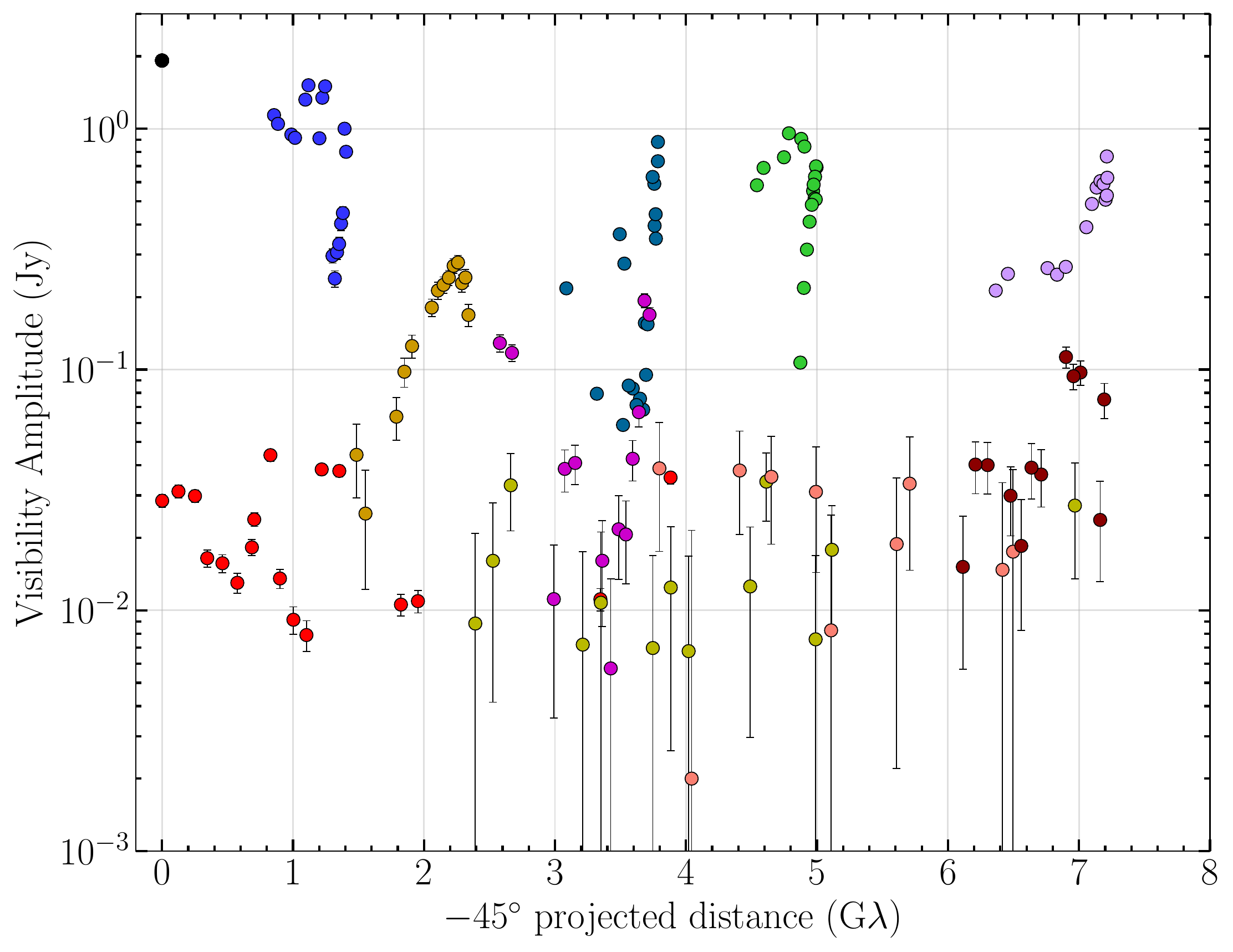}
      \end{subfigure}
    \end{minipage}
    \caption{\ac{vlbi} image reconstruction and underlying visibility amplitude data reproduced from \citet{2021Janssen}. The results are from EHT observations of Centaurus~A in 2017 at 228\,GHz. The image shown in the left panel was reconstructed with the \texttt{eht-imaging} regularized maximum likelihood method. The right panel shows the visibility amplitudes as a function of \uv{}-distance projected along the jet direction (\textit{top}) and perpendicular (\textit{bottom}) to the jet direction.}\label{fig:im}
\end{figure}

When imaging \ac{vlbi} data, one tries to reconstruct the sky brightness distribution that fits the data best and satisfies additional assumptions imposed to deal with the ill-posed imaging problem of incompletely sampled Fourier data from a sparse array.
Strictly speaking, ``imaging'' refers to obtaining a pixel-based model (convolved with a resolving beam) through inverse modeling or forward modeling \citep{2019ApJ...875L...4E}.
Inverse modeling techniques like the CLEAN algorithm \citep{clean1,clean2} gradually reconstruct a model in the image domain of the inverse Fourier transform of the visibilities.
There are a couple of different CLEAN implementations in different software packages that are worth mentioning. The original H\"ogbom CLEAN method \citep{clean1} uses iterations of delta image model components convolved with the point spread function (PSF) that are subtracted from the ``dirty'' image of the inverse Fourier transform. The process converges when the residual map consists only of noise. The collection of delta components form the final image model. When the residual image is added and the delta components are convolved with the \ac{vlbi} resolving beam, the final image is obtained. Knowledge about interferometric images is needed to identify ``false'' sidelobe emission regions and to put boxes (``CLEAN windows''), which limit the locations in which CLEAN can put delta components: the emission peaks within the CLEAN windows.
\citet{clean2} invented a faster variant of the H\"ogbom CLEAN for large images.
Clark CLEAN operates in two cycles; the minor cycle works like the H\"ogbom CLEAN, except that only the inner ``beam patch'' of the PSF is used. In major cycles, the collection of delta components are then subtracted with the full PSF. Multi-scale CLEAN \citep[MSC,][]{2008ISTSP...2..793C} allows the use of extended Gaussian components for the model next to the delta components. This proved to be useful in accurately deconvolving extended / diffuse emission.

These techniques used to average the frequency bandwidth into a single channel to improve \ac{sn}. However, the development of multi-frequency synthesis \citep[MFS,][]{1990MNRAS.246..490C,1994A&AS..108..585S,2005ASPC..340..608L} allowed the gridding of visibilities measured across multiple frequencies onto the same $(u,v)$ plane, thus improving the fidelity of the resultant image. With the widening bandwidths of modern instruments, the in-band source spectral indices limited the deconvolution accuracy when using MFS. To solve this, the multi-term multi-frequency synthesis \citep[MT-MFS,][]{2011A&A...532A..71R} algorithm models the frequency dependence of the source using a Taylor expansion. These various algorithms can be used in combination with each other depending on the observational configuration and source structures. For example, the Multi-scale multi-frequency synthesis \citep[MS-MFS,][]{2011A&A...532A..71R} combines MSC with MT-MFS.  

Forward modeling techniques such as the maximum entropy method \citep{1985A&A...143...77C} fit the Fourier transform of an image to the visibility data. Additional assumptions can be incorporated as regularizers. Fitting of geometric forms such as two-dimensional Gaussian components are also done with the forward modeling approach in the visibility domain. \Cref{fig:im} shows an example \ac{vlbi} image reconstruction together with the underlying visibility amplitude measurements. Projected along the jet direction, the visibility amplitudes are indicative of a smooth Gaussian-like structure. Perpendicular to the jet direction, bounding amplitudes are indicative of strong intensity gradients across the transverse jet profile.

While the \ac{vlbi} scientific analysis is usually performed in the image domain, it is also possible to compare physics-based theoretical source models directly to the visibility measurements.
The direct data-model comparison method will likely become more prominent in the future with the availability of increasingly advanced \ac{agn} jet models\citep[e.g., see][for a recent review]{2022arXiv220112608M} and machine learning methods \citep{2020A&A...636A..94V, 2021arXiv211007185Y}.
For the analysis of M87* and Sgr~A* EHT observations, horizon-scale simulations \citep{2019ApJ...875L...4E} are directly compared to the visibility data with (proprietary) software packages such as \texttt{GENA} \citep{2019A&A...629A...4F} and \texttt{THEMIS} \citep{2020ApJ...897..139B}.

Specific combinations of the baseline-based visibilities can be formed as ``closure quantities'', where station-based data errors drop out \citep{1958MNRAS.118..276J, 2020Blackburn, ClosureTraces2020, 2022PhRvD.105d3019T}. For example, the sum of baseline phases in a closed triangle form a robust closure phase.
While uncorrected residual gain errors negatively impact reconstructed images when the full set of visibilities are utilized, with only closure quantities, one does not make use of the full information content of the data.
All imaging software packages listed below (\Cref{sec:imsoftware}) are able to perform \textit{self-calibration}, where, for sufficiently bright sources, within given solution intervals, per-station phase and/or amplitude gains can be found with a least-squares solver based on a comparison between the observed data and model visibilities.
Typically, an initial self-calibration is performed based on a simple starting model or an initial image reconstruction where only closure quantities have been used.
Subsequently, one can iteratively image and self-calibrate until the gain solutions are converged.
This method of solving for the observed source structure together with calibration gains is referred to as ``hybrid-mapping'' \citep{1978ApJ...223...25R, 1979AJ.....84.1122C}.

All imagers listed below can make use of the \texttt{UVFITS} visibility data format as input and produce \texttt{FITS} images as output.
Full Stokes polarimetric imaging can be done with each imager. For details, the reader is referred to \citep[][and Mart\'i-Vidal et al., in prep.]{2021A&A...646A..52M}.

\subsection{Software implementations}
\label{sec:imsoftware}

\subsubsection{AIPS}

The \ac{aips} \texttt{IMAGR} task implements the Clark CLEAN method.

\subsubsection{CASA}

The \ac{casa} \texttt{tclean} task implements the H\"ogbom, Clark, MSC, and MS/MT-MFS deconvolver options for CLEAN.
Next to \texttt{UVFITS} input, the \ac{casa} imager works with its native \texttt{MS} visibility and metadata file format.

\ac{casa} is commonly used for spectral-line and wide-field observations, for which the corresponding special \texttt{tclean} features are described in \Cref{sec:line} and \Cref{sec:wide}, respectively.

\subsubsection{Difmap}

\href{ftp://ftp.astro.caltech.edu/pub/difmap/difmap.html}{\texttt{Difmap}} \citep{1997ASPC..125...77S}\footnote{\url{ftp://ftp.astro.caltech.edu/pub/difmap/difmap.html}.} is a widely used \ac{vlbi} imaging and geometric model fitting software owing to its computational efficiency, intuitive user interface, and convenient data plotting plus flagging procedures.
\texttt{Difmap} implements H\"ogbom CLEAN.

\subsubsection{WSClean}

\href{https://gitlab.com/aroffringa/wsclean}{\texttt{WSClean}} \citep{2014MNRAS.444..606O}\footnote{\url{https://gitlab.com/aroffringa/wsclean}.} has been developed with the primary goal of computationally efficient wide-field imaging, for which the corresponding special software features are descried in \Cref{sec:wide}.
The latest version of the software can use the Clark CLEAN implementation and reconstruct MS-MFS images computationally more efficient than \texttt{tclean} \citep{2017MNRAS.471..301O}.

\texttt{WSClean} uses the \texttt{MS} file format. \texttt{UVFITS} input can easily be converted with the \ac{casa} \texttt{importuvfits} task.

\subsubsection{eht-imaging}
\label{sec:ehtim}

The \href{https://github.com/achael/eht-imaging}{\texttt{eht-imaging}} software package \citep{ehim1, ehim2}\footnote{\url{https://github.com/achael/eht-imaging}.} is a collection of \texttt{Python} modules for \ac{vlbi} data analysis, with a focus on regularized maximum likelihood (RML) imaging. It includes routines for detailed mock data simulation, gain and polarimetric calibration of visibilities, image and data plotting and statistical analysis, and RML image reconstruction as well as geometric model fitting from visibilities and closure products (including full-Stokes and multi-frequency data).

\subsubsection{SMILI}

\href{https://github.com/astrosmili/smili}{\texttt{SMILI}} \citep{smili1,smili2}\footnote{\url{https://github.com/astrosmili/smili}.} is a collection of sparse sampling RML forward modeling libraries, similar to \texttt{eht-imaging}.
\texttt{SMILI} can be used through a \texttt{Python} interface.

\subsubsection{UVMULTIFIT}

\href{http://mural.uv.es/imarvi/docums/uvmultifit}{\texttt{UVMULTIFIT}} \cite{UVMULTIFIT}\footnote{\url{http://mural.uv.es/imarvi/docums/uvmultifit}.} is a flexible based geometric model fitting library that is integrated in the \ac{casa} ecosystem. 

\subsubsection{Comrade and DPI}

The \texttt{Julia}-based \citep{Julia} \texttt{Comrade} \citep{2022ApJ...925..122T} and \texttt{Python}-based \texttt{Deep Probabilistic Imaging/Inference} \citep[\texttt{DPI/$\alpha$-DPI},][]{DPI1, DPI2} frameworks are versatile geometric model fitting tools, which are used by the EHT \citep{eht-SgrAiii}.

\section{Advanced scientific applications}
\label{sec:special}

\subsection{Polarization calibration}
\label{sec:pol}

For polarization considerations, we consider the sky signal to be split into independent right-handed and left-handed circular polarizations (RCP and LCP, respectively). Such a signal split is commonly obtained in \ac{vlbi} receiving systems by placing quarter-waveplates in front of linear polarization feeds.
Advanced calibration methods, that have traditionally been developed for the inclusion of ALMA in \ac{vlbi} experiments \citep{2016A&A...587A.143M}, enable routine conversions of recorded linear polarization signals to circular ones during correlation (\Cref{sec:corr}).

\subsubsection{Additional polarization signal stabilization steps}

The first step is to properly align the RCP and LCP signals.
The geometric feed rotation angle evolution between the RCP and LCP data (commonly referred to as parallactic angle rotation) depends on a telescope's focus and mount configuration \citep[e.g., see Appendix C of][]{rpicard} and is usually taken out when the data are fringe-fitted (\Cref{sec:signal}).\footnote{Note that the telescope focus and mount configuration information are not always correctly specified in the visibility data files produced by the correlators. For all software packages described here, there are methods to overwrite this information manually.}
When all RCP and LCP fringe solutions are (re-)referenced to a common station in a stable system, a global cross-hand R-L delay will be present in the data.
This is the instrumental delay between the RCP and LCP signal chains of the chosen reference station, that can be determined by fringe-fitting the RL and LR visibilities.
The R-L phase is similarly affected by instrumental effects of the reference station. The true absolute R-L phase sets the orientation of the electric vector polarization angle (EVPA) of the polarized radio emission, which has to be bootstrapped from other (e.g., single-dish) observations.

If the fringe-fitting is done separately for RCP and LCP, the R-L alignment can be done as final calibration step. If polarizations are combined for the fringe-fitting, the alignment should be done as part of the instrumental calibration steps (\Cref{fig:calibrationrFlow}).

\subsubsection{Solving for polarization leakage effects}

Subsequently, the mutual leakage of signals between the RCP and LCP signal paths at every station should be calibrated. If left uncorrected, these leakages or ``$D$-terms'' will impose small errors in total intensity data and add polarization signals that are purely instrumental (i.e., Stokes $\mathcal{Q}$ and $\mathcal{U}$ signals can occur for unpolarized sources).
$D$-terms can have a frequency dependence, are usually stable in time, and small enough that the signal stabilization calibration does not have to be revised with leakage-corrected visibilities.
Before leakages can be determined, a good image (\Cref{sec:image}) of the observed source is required.
Firstly, to have gain errors removed by self-calibration and secondly, because solving for $D$-terms for resolved sources with complex polarization structures is not trivial \citep{2021A&A...646A..52M}.
For a thorough and recent overview of $D$-term calibration methods and software implementations, the reader is referred to \citep[][and Mart\'i-Vidal et al., in prep.]{2021A&A...646A..52M}.

\subsubsection{Circular polarization}

For Stokes $\mathcal{V}$ studies of \ac{vlbi} data in a circular polarization basis, self-calibration solutions obtained from the RR and LL visibilities are needed to estimate the relative R/L amplitude gains of every antenna in the array. The problem is that an intrinsic Stokes $\mathcal{V}$ source polarization will confound the true residual R/L gains. A common remedy is to plan an observation where R/L gains from many observed sources based on Stokes $\mathcal{I}$ images can be obtained. Under the assumption that $\mathcal{V}=0$ on average for the sample, as there is no intrinsic preference for positive or negative intrinsic circular polarization, the mean R/L gains should be applied to the data \citep{1999AJ....118.1942H}.

\subsection{Spectral line observations}
\label{sec:line}

For telescopes equipped with spectrometers and a sufficient frequency resolution set at the correlator, emission and absorption features from spectral lines can be studied.

\subsubsection{Spectral line signal stabilization}

For spectral line observations, a few special considerations and adjustments of certain steps have to be taken into account for the signal stabilization \citep[e.g.,][]{2004Goddi}.
Fringe-fit delay solutions and bandpass corrections should be obtained only on continuum sources.
Depending on the reference frame used during correlation, Doppler shifts of spectral lines have to be corrected to keep the position of the line constant with respect to the correlated frequency channels after delays and the bandpass have been corrected. The center of the Earth is usually chosen as reference point during correlation and small corrections that are common for all antennas due to the Earth's orbit around the Sun relative to the observed source are applied when shifting to the Local Standard of Rest. Such effects become increasingly relevant for longer observations.
Fringe rates can be determined on the science target, selecting one or a few channels with strong line emission.
Strong line emission can also be used for phase-referencing.

The \ac{aips} and \ac{casa} fringe-fitting tasks are able to handle spectral line observations. The Doppler shifts can be corrected with the combined \texttt{SETJY} and \texttt{CVEL} tasks in \ac{aips} and the \texttt{mstransform} task in \ac{casa}.
The \ac{casa} imager is able to reconstruct spectral cubes (\Cref{sec:image}).

\subsubsection{Template spectrum flux density calibration}

Bright lines will show up in telescope's auto-correlation spectra, exceeding the atmospheric noise. A \textit{template spectrum} can be obtained from the measured bandpass-corrected total power response of the line from a sensitive antenna in the \ac{vlbi} array. Amplitude gains in the form of time-dependent \ac{sefd}s can then be obtained without using system temperatures (\Cref{sec:flux}) by fitting the total power spectral of all antennas to the template \citep[][Lecture 12]{vlbiTA}.

This calibration method is implemented in the \ac{aips} task \texttt{ACFIT}. 

\subsubsection{Spectral line imaging}

For spectral lines, the self-calibration is best performed with the peak line emission channel(s) image. Subsequently, spectral cube images can be made from the channels/velocities where line emission is present.
If necessary, \ac{casa} \texttt{tclean} can perform Doppler tracking to keep the positions of spectral lines stable in the Kinematic Local Standard of Rest.
For an example of a particularly extensive spectral line \ac{vlbi} imaging work, the reader is referred to \citet{2010Matthews}.

\subsection{Wide-field VLBI}
\label{sec:wide}

In wide-field \ac{vlbi} experiments, multiple sources are observed within single pointing fields that cover the full or a signification fraction of the participating telescopes' primary beams.\footnote{See \citet{2004evn..conf..273S} for the primary beam of a heterogeneous interferometer array.}

\subsubsection{Wide-field correlation}

For us to observe sources across the entire primary beam using VLBI, we require correlation to be conducted at a very high time (typically millisecond) and frequency ($\sim$kHz) resolutions in order to restrain smearing at the edge of the primary beam. However, the result is that a single large and often unwieldy data-set is produced which makes calibration, imaging and subsequent analysis difficult without significant computing resources. This method has been used for many of the early wide-field VLBI experiments \citep[e.g.,][]{2013A&A...550A..68C} but the computational complexity and correlation resources to map the entire primary beam made wide-field VLBI observations an unattractive concept. 

An alternative correlation method for wide-field VLBI was developed with the \texttt{DiFX} software correlator \citep{2011A&A...526A.140M} and has been implemented in the JIVE \texttt{SFXC} correlator \citep{2015ExA....39..259K}. Named ``multiple simultaneous phase centre observing'', this uses a two-step correlation approach. Firstly, an internal wide-field correlation produces data at the required time and frequency resolution to restrain smearing to acceptable levels at the edge of the primary beam. These data are copied and phase rotated to multiple positions across the primary beam \ac{fov}, which retains high precision astrometry as an internal geometric delay model within the correlator applies the required phase rotation. Each of these data-sets, which are centered at different positions, are then averaged to coarse time and frequency resolution, producing multiple pencil-beam data sets across the primary beam. These pencil-beams can be arranged to cover multiple sources discovered in accompanying low-resolution surveys or be mosaicked to cover a region of interest or even the whole primary beam \ac{fov}. The result from this mode of correlation is multiple small and manageable data-sets (often $\sim$GB-sized), one for each position.



\subsubsection{Wide-field signal stabilization}

Signal stabilization steps (\Cref{sec:signal}) can be applied to wide-field observations as usual, by applying the calibration solutions to the data from all phase centers. Calibrator sources may be found within the phase centers as in-beam calibrators or from another area of the sky covered in a different pointing direction.

\subsubsection{Wide-field imaging: primary beam correction, the $w$ term, and multi-source phase self calibration}

When the beams of the participating telescopes are well determined, primary beam corrections can be employed across the \ac{fov}. For homogeneous arrays, such as the VLBA, the primary beam effectively adjusts the flux densities with distance from the pointing centre as each baseline sees the same apparent sky brightness distribution that is modified by the primary beam power envelope\footnote{We note that this is not strictly the case outside the main primary beam lobe, especially for non-equatorial mounts, where the different parallactic angles at each antenna results in different side-lobe structure.}. 

However, most VLBI arrays contain heterogeneous elements, meaning that the primary beam has significant effect even within the main lobe of the primary beam. This manifests in the data as a direction-dependent and antenna-independent gain error [see Radcliffe et. al. in prep.]. Corrections for this effect must be conducted in the \uv-plane and can be achieved in two ways. The first is to produce a direction-independent gain table with the corrections corresponding to the phase centre position (implemented through the task \texttt{CLVLB} in \texttt{AIPS} and internally within the \texttt{VPIPE} pipeline). This is computationally inexpensive, can account for the frequency and time dependence of the primary beam, and can be applied together with the phase-referencing corrections.

The second method is to correct during the gridding of the image reconstruction using the convolution of the complex conjugates of the primary beam voltage responses. This is more computationally complex but any residual errors are only dependent upon the accuracy of the primary beam models. Prototypes have been implemented in \texttt{WSClean} through the image-domain gridder (IDG) \citep{vandertol2019} through the application of a direction-dependent diagonal gain correction [see Radcliffe et al. in prep.]. 


This correction is applied together with corrections for a non-negligible $w$ term \citep{TMS} that causes increasingly larger errors in images with distance from the phase center due to sky curvature and non-coplanar baselines. Due to the $w$ term, the Fourier-transform between the visibilities and the sky brightness distribution is no longer two-dimensional. A full three-dimensional FFT would result in strong aliasing along the $w$ direction and a direct transform along $w$ is computationally too expensive for a cube that is mostly devoid of emission.

The $w$ term effects are typically corrected with one of the following methods:
\begin{itemize}
     \item Faceting \citep{1992A&A...261..353C} performs a 2D Fourier-transform for a number of different phase centers across the \ac{fov}. The resultant pieces of image facets are small enough that the 2D FFT approximation works. The facets are stitched together and deconvolved.
     \item $w$-projection \citep{2008ISTSP...2..647C} uses the fact that the visibility $\mathcal{V}$ as a function of $w$ can be calculated from $\mathcal{V}$ at $w=0$ through a convolution operation: $\mathcal{V}(u,v,w) = \widetilde{G}(u,v,w) * \mathcal{V}(u,v,w=0)$. Here, $\widetilde{G}(u,v,w)$ is the Fourier transform of ${G}(u,v,w) = \exp\left[-2 \pi i w \left(\sqrt{1-l^2-m^2} - 1 \right)\right]$. This method is faster than faceting because the visibilities need to be gridded only once.
     \item $w$-stacking \citep{2014MNRAS.444..606O} uses a $w$-dependent grid which is FFT'ed for each gridded $w$ value and phase-shifted by ${G}^{-1}(u,v,w)$. For imaging, all grids are summed with appropriate scaling factors \citep[see][for details]{2014MNRAS.444..606O}. Compared to $w$-projection, $w$-stacking is faster when the visibility gridding is computationally more expensive than the FFT computations.
 \end{itemize}
%
\texttt{WSClean} implements the $w$-stacking method and $\texttt{tclean}$ implements the $w$-projection algorithm and can do faceting.

The phase stability of VLBI arrays is worse than connected short-baseline arrays due to the differing atmospheric paths between antenna and source. This means that self-calibration (\Cref{sec:image}) is essential in achieving low noise and high dynamic ranges. Self-calibration requires a source with sufficient S/N to estimate the gain corrections along the line-of-sight to the target field. However, for VLBI observations, its lack of sensitivity to large scale, diffuse emission means that the total flux available for self-calibration of a single source can be insufficient, restricting VLBI observations to those fields with in-beam, or nearby phase calibrators.

However, for wide-field VLBI, the targeting of multiple sources allows self-calibration to proceed, even when each individual target is faint, by using the combined flux density of the detected sources \citep{middelberg2013, 2016A&A...587A..85R}. This technique, called ``multi-source self-calibration'', images each source detected after phase referencing and divides the visibilities are divided by the model to produce a point source at the phase centre. These data sets are stacked in the \uv-plane and self-calibration solutions obtained using the combined S/N of all target sources. This technique is currently being developed into a fully directional-dependent technique through the derivation of self-calibration solutions along different line-of-sights across the primary beam [Harth et al. in prep.].


\section{Synthetic data}
\label{sec:synth}

Forward-modeling methods, where the \uv-coverage, thermal noise and simple telescope gain errors are taken into account to create simulated/fake visibility data based on a model sky brightness distribution, have been available for a while (e.g., the \texttt{DTSIM} \ac{aips} task, \texttt{simobserve} in \ac{casa}, and MeqTrees \citep{2010A&A...524A..61N}\footnote{\url{http://meqtrees.net}.}).
Recently, very accurate visibility simulations have been developed: The data simulation toolkit within \texttt{eht-imaging} and the \ac{symba} \citep{SYMBA}\footnote{\url{https://bitbucket.org/M_Janssen/symba}.}.
Both methods make use of sophisticated data corruption models, are able to include the effects of interstellar scattering, can use arbitrary source models defined in text files or images from \texttt{HDF5} and \texttt{FITS} data, and use the exact \uv{}-coverage from (observational) \texttt{UVFITS} input.
\ac{symba} and \texttt{eht-imaging} are routinely used by the EHT to generating synthetic data for dedicated tests to answer specific scientific questions (e.g., about the robustness of specific features in reconstructed images), optimize calibration procedures, produce imaging parameter surveys, create test suites for software packages, and identify how to optimally upgrade \ac{vlbi} arrays \citep[e.g.,][]{EHT2019I, 2019ApJ...875L...4E, 2019ApJ...875L...6E, 2020ApJ...901...67W, 2021A&A...650A..56R, 2021NatAs...5.1017J}.

\begin{figure}[t]
\centering
\includegraphics[width=0.95\columnwidth]{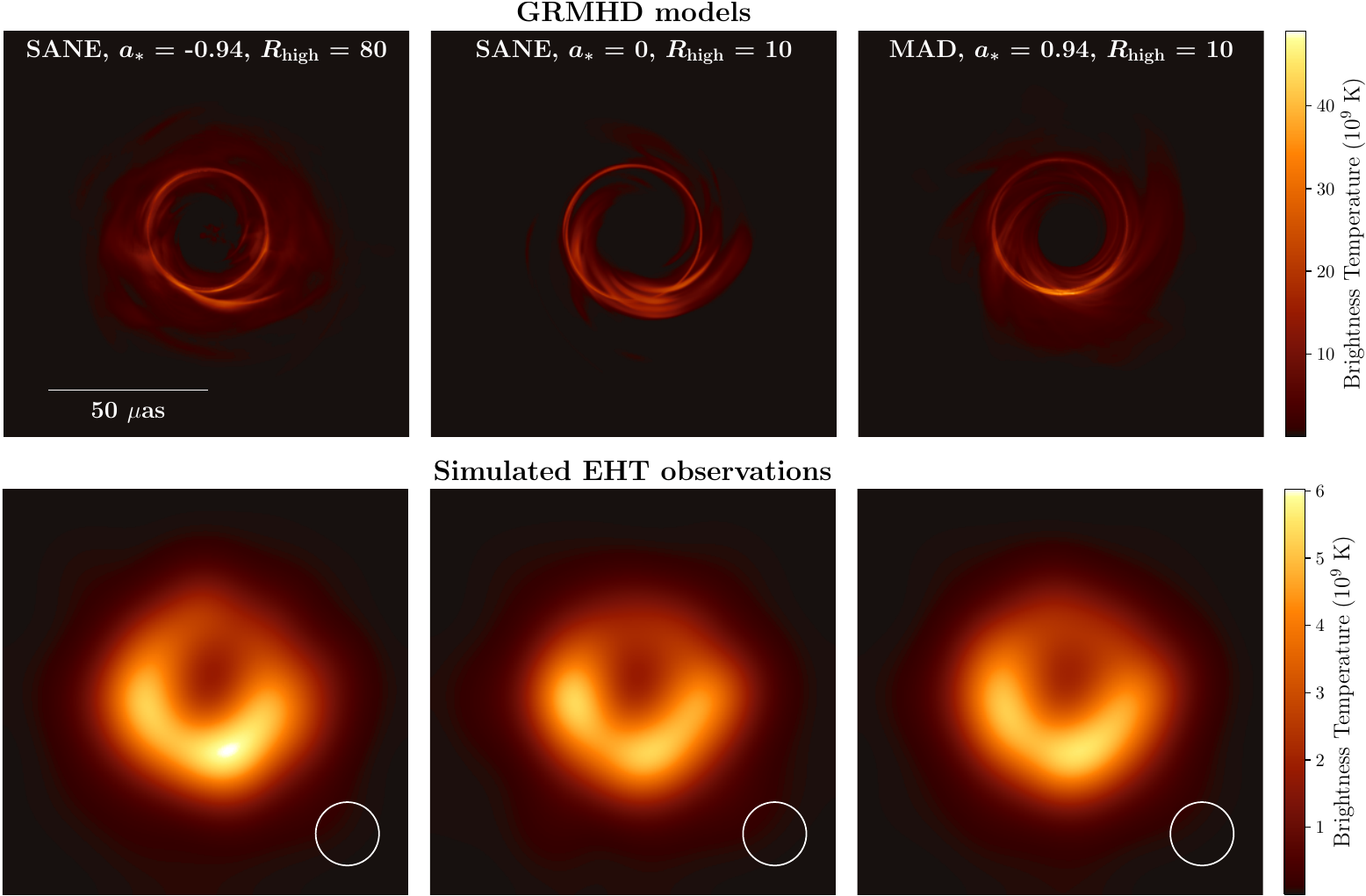}
\caption{Synthetic EHT observations of M87*. \textit{Top}: Ray-traced \ac{grmhd} simulations \citep{2019ApJ...875L...5E} from the EHT library with different black hole spins, electron temperature parameters, and magnetic field states (``MAD'' versus ``SANE'').
    \textit{Bottom}: \ac{symba} synthetic observations corresponding to these models. Reproduced from \citet{EHT2019I}.}
\label{fig:symba-eht}
\end{figure}

The versatile \texttt{eht-imaging} package bootstraps data corruption effects from a calibrated \texttt{UVFITS} file of observational data and produces a \texttt{UVFITS} output.
The \ac{casa}-based \ac{symba} software produces both \texttt{UVFITS} and \ac{ms} output. Data corruptions effects are simulated from first principles with MeqSilhouette to emulate the correlator output in a real observation \citep{2017Blecher,2022Natarajan}\footnote{\url{https://github.com/rdeane/MeqSilhouette}.}.
For example, physical parameters for atmospheric or antenna-pointing models can be adjusted and their effects studied.
The suite of possible MeqSilhouette data corruption effects is continuously being expanded; Faraday rotation effects from the interstellar medium and atmospheric effects from the ionosphere will soon be added for example.
\ac{symba} passes the MeqSilhouette output to \ac{rpicard}, which results in science-ready calibrated synthetic data akin to that from a real observation. Work is ongoing to convert \ac{symba} to simulate low frequency VLBI and enable feasability studies of new instruments such as SKA-VLBI.

The ability to simulate \ac{vlbi} data based on known ground-truth models allow non-experts to better understand the fundamentals of \ac{vlbi} data.
For example, by looking at the amplitude and phase signatures of simple geometric models or by studying how specific calibration errors manifest themselves in the data.
Examples of synthetic EHT observations based on state-of-the-art theoretical models are shown in \Cref{fig:symba-eht}.

Next to the full forward-modeling software packages, there are also dedicated \ac{vlbi} observing simulators, such as the \texttt{Very long baseline interferometry Network SIMulator (VNSIM)} \citep{2019JKAS...52..207Z}\footnote{\url{https://github.com/ZhenZHAO/VNSIM}.}.
These tools assist users and engineers with the design and scheduling of \ac{vlbi} experiments, as well as the evaluation of the array performance under various conditions.


\section{Summary}
\label{sec:summary}

In this work we have presented current open-source \ac{vlbi} methods and their implementations in various software.
\Cref{tab:softwares} provides and overview of the major software packages' capabilities. Several of these packages can be easily obtained through the \href{https://kernsuite.info}{KERN repository} \citep{molenaar2018kern}\footnote{\url{https://kernsuite.info}.} alongside many additional radio astronomical software libraries.

\begin{table}[H]
\caption{Major \ac{vlbi} software packages and their capabilities. \texttt{AIPS} and \texttt{Difmap} are still supported by their originators; requested features and bug fixes can be implemented. All other packages are actively being developed and supported by larger communities and institutions.}
\centering
\begin{tabular}{p{2.1cm}p{3.9cm}p{7.8cm}}
\toprule
\textbf{Name}	& \textbf{Capabilities}	& \textbf{Links and references}\\
\midrule
\texttt{DiFX}		& Correlation			&  \url{https://ascl.net/1102.024},\;\;\newline\url{https://svn.atnf.csiro.au/difx},\;\;\cite{2007PASP..119..318D,2011PASP..123..275D} \\ \\
\texttt{SFXC}		& Correlation			&  \url{https://svn.astron.nl/sfxc},\;\;\cite{2015ExA....39..259K}\\ \\ 
\texttt{AIPS}, \texttt{ParselTongue}        & Signal stabilization, flux density calibration, imaging, synthetic data generation  & \url{https://ascl.net/9911.003},\;\;\url{https://ascl.net/1208.020},\newline\url{http://www.aips.nrao.edu},\;\;\newline\url{https://www.jive.eu/jivewiki/doku.php?id=parseltongue:parseltongue},\;\;\citep{2003Greisen, 2006Kettenis} \\ \\
\texttt{CASA}\newline\newline \hspace*{0.1cm}- \texttt{VPIPE}\newline\newline \hspace*{0.1cm}- \ac{rpicard}       & Signal stabilization, flux density calibration, imaging & \url{https://ascl.net/1107.013},\;\;\url{https://casa.nrao.edu}\;\;\newline\newline\url{https://github.com/jradcliffe5/VLBI_pipeline}\newline\newline\url{https://ascl.net/1905.015},\;\;\newline\url{https://bitbucket.org/M_Janssen/picard},\;\;\newline\url{https://hub.docker.com/r/mjanssen2308/casavlbi},\;\;\newline\cite{2007ASPC..376..127M, 2018evn..confE..80J,rpicard} \\ \\
\texttt{(EHT-)HOPS} & Signal stabilization, flux denisity calibration & \mbox{\url{https://www.haystack.mit.edu/tech/vlbi/hops.html}},\newline\url{https://github.com/sao-eht/eat},\;\;\cite{2004RaSc...39.1007W, 2019ApJ...882...23B} \\ \\
\texttt{Difmap} & Imaging, model fitting  & \url{https://ascl.net/1103.001},\;\;\newline\url{ftp://ftp.astro.caltech.edu/pub/difmap/difmap.html},\;\;\citep{1997ASPC..125...77S} \\ \\ 
\texttt{eht-imaging} & Imaging, model fitting, synthetic data generation & \url{https://ascl.net/1904.004},\;\;\newline\url{https://github.com/achael/eht-imaging},\;\;\citep{ehim1, ehim2} \\ \\
\texttt{SMILI} & Imaging  & \url{https://ascl.net/1904.005},\;\;\newline\url{https://github.com/astrosmili/smili},\;\;\citep{smili1,smili2}\\ \\
\texttt{WSClean} & Imaging & \url{https://ascl.net/1408.023},\;\;\newline\url{https://gitlab.com/aroffringa/wsclean},\;\;\citep{2014MNRAS.444..606O} \\ \\
\texttt{UV-\newline{MULTIFIT}} & Geometric model fitting & \url{https://ascl.net/1402.017},\;\;\newline\url{https://launchpad.net/uvmultifit},\;\;\citep{UVMULTIFIT} \\ \\
\ac{symba} & Synthetic data generation & \url{https://bitbucket.org/M_Janssen/symba},\;\;\citep{SYMBA} \\
\bottomrule
\label{tab:softwares}
\end{tabular}
\end{table}

\acknowledgments{The authors thank Ciriaco Goddi, Freek Roelofs, and Ilse van Bemmel for their valuable comments to the manuscript. We also thank the anonymous referees for helpful comments that improved the paper.} 



\appendixtitles{no} 
\appendixsections{multiple} 
\newpage
\appendix

\section{Telescope baseband data transport to the VLBI correlator}
\label{sec:corr-data}

Baseband data from all stations need to be available simultaneously for correlation. One option is real-time streaming to the VLBI correlator. This is an observing mode in e-MERLIN \citep{MERLIN}, e-EVN, and future VLBA after ongoing upgrades. 

Real-time correlation is possible if the network paths from the VLBI telescopes (connected at typically 1G/10G/100G) to the correlator (typically 10G/100G) offer sufficient capacity and can maintain real-time rates reliably with an acceptably low data loss fraction. On paths that cross the Internet this can be challenging. A drawback is also that re-correlations with different settings are not possible, unless the baseband data are stored.

In most VLBI experiments the baseband data are captured at the telescopes onto a recorder such as  FlexBuff\footnote{JIVE FlexBuff RAID/JBOD recorders, $\ge4$~Gbps, scalable, stationary, \url{https://www.jive.eu/technical-operations-rd-group}.},  Mark~6\footnote{Conduant/MIT Haystack Mark~6, max. 32~Gbps, shippable, \url{https://www.haystack.mit.edu/mark-6-vlbi-data-system/}.}, or Octadisk2\footnote{Elecs/NAOJ Octadisk2, max. 32~Gbps, shippable, \url{https://www.elecs.co.jp/product/removable_storage.html}.}.
The recording media can be physically shipped by courier mail. The proprietary form factor of the media requires a corresponding playback unit at the correlator. 

Alternatively, baseband data files can be transferred at slower pace over the Internet to the correlator.
Popular high speed data transfer software for this purpose are JIVE {\em jive5ab}\footnote{\url{https://github.com/jive-vlbi/jive5ab}.} and {\em etc/etd}\footnote{\url{https://github.com/jive-vlbi/etransfer}.}, {\em Tsunami UDP} transfer\footnote{\url{https://tsunami-udp.sourceforge.net}.}, and Globus {\em GridFTP}\footnote{\url{https://www.globus.org/}.}. Transfer with slower {\em scp} or {\em rsync} is also used as a fallback.
In an ``e-shipping'' mode, the transferred data are stored on a cluster or cloud file system until later correlation.
In contrast, real-time correlation of the transferred data is commonly referred to as ``e-VLBI''.

\section{Data flagging}
\label{sec:flagging}

\begin{figure}
\centering
\includegraphics[width=0.9\columnwidth]{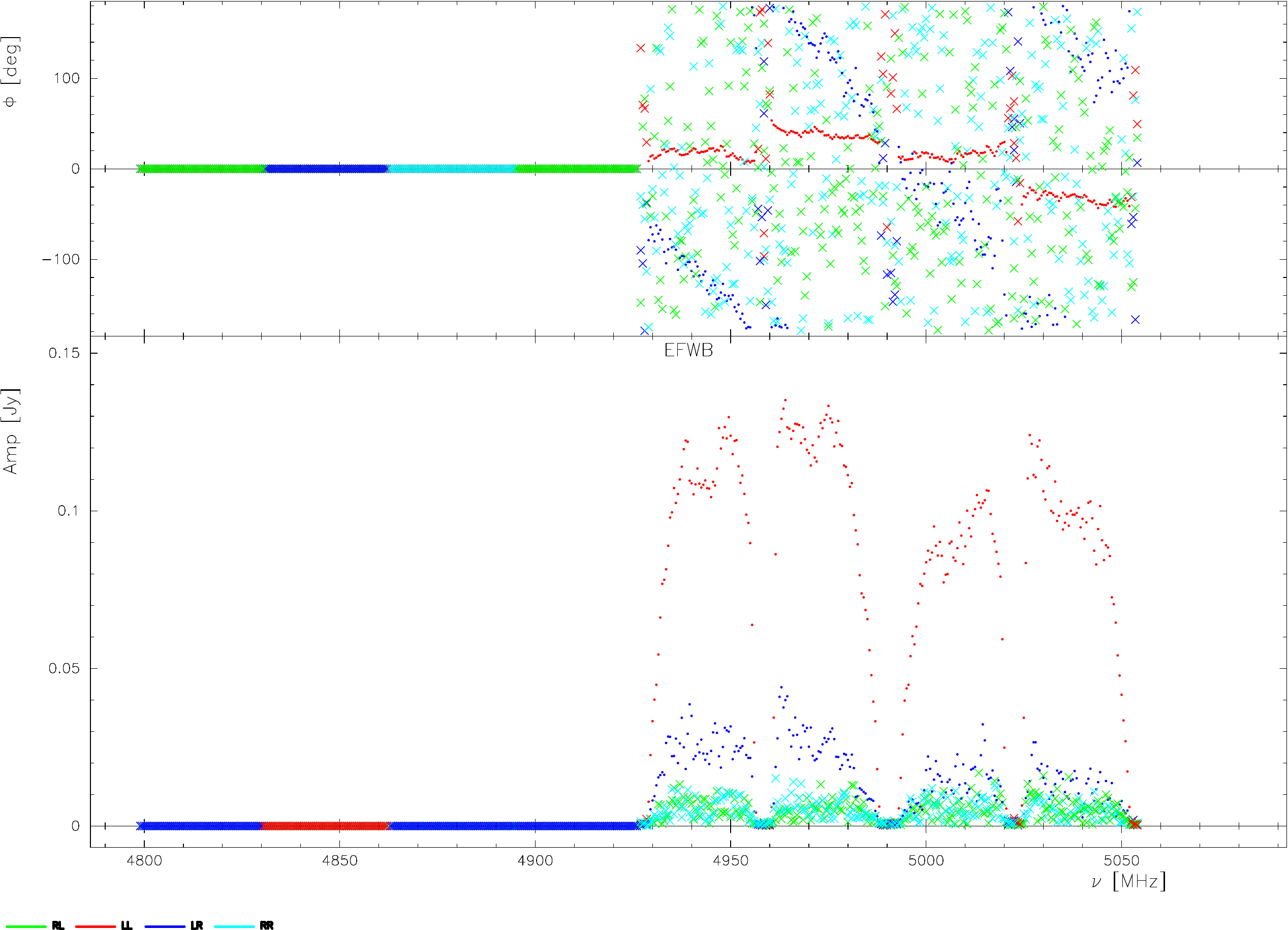}
\includegraphics[width=0.9\columnwidth]{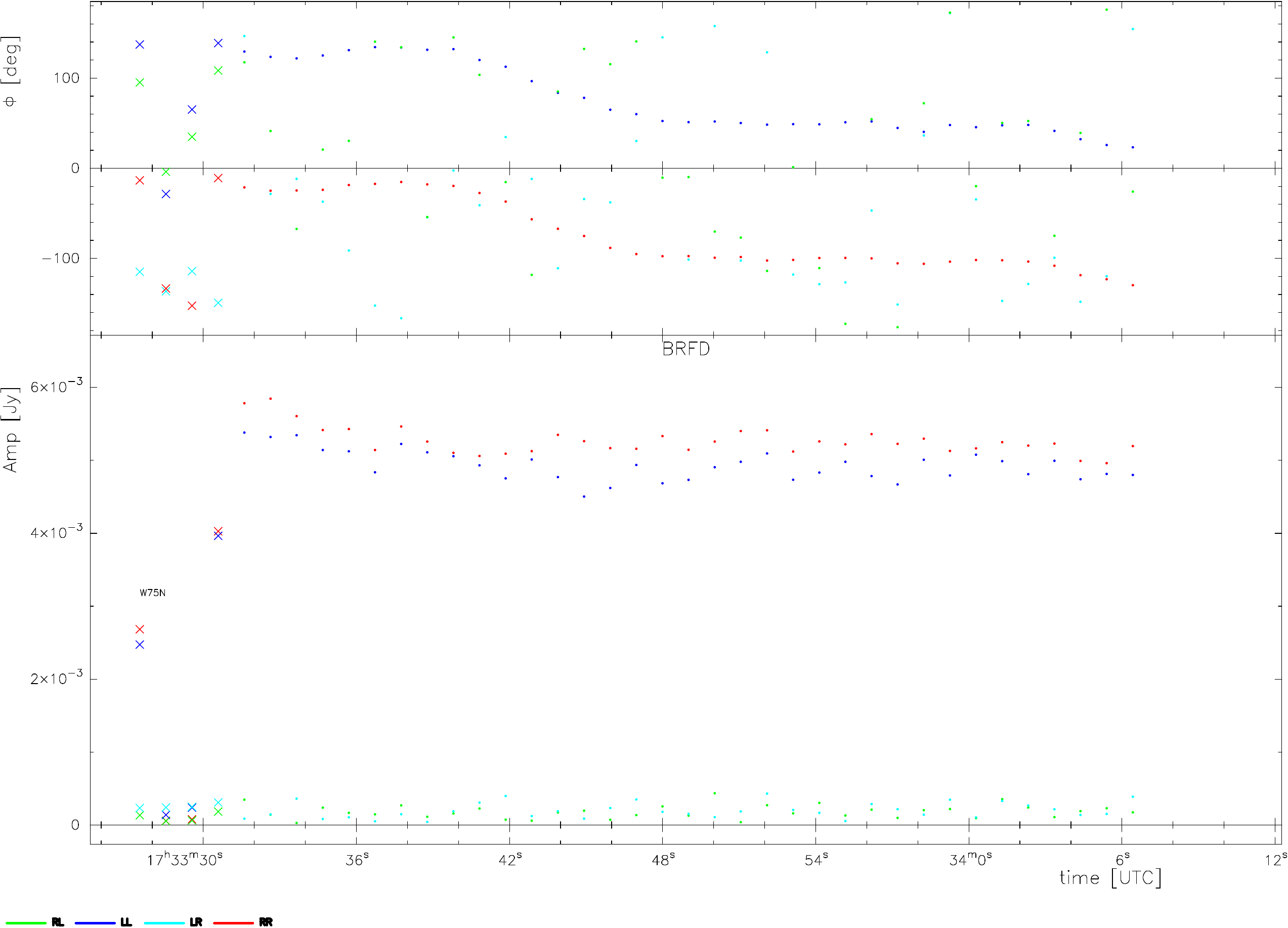}
\caption{Examples of flagged bad data indicated with crosses. These plots are produced automatically with \href{https://github.com/haavee/jiveplot}{jiveplot} by \ac{rpicard} and show visibility amplitudes and phases per baseline and per scan, color-coded by polarization. \textit{Top}: Scan-averaged EVN data on the Effelsberg-Westerbork baseline. The Effelsberg RCP channel and first four spectral windows contain no signal and have been flagged. \textit{Bottom}: Frequency-averaged VLBA data on the Brewster-Hancock baseline. Brewster has been a few seconds late on source, resulting in bad phases and amplitudes at the start of the scan.}
\label{fig:flag}
\end{figure}

The flagging (removal) of ``bad data'' is common practice in interferometry. With enough antennas in the array, one can be rather liberal with the data flagging. \ac{vlbi} arrays, however, are often sparse.
Care must be taken to distinguish bad data from correctable calibration errors and measurements with low \ac{sn}.
Common uncorrectable problems are issues along the telescope data recoding path that may render data at specific frequency channels or polarizations unusable, telescopes that are late on source or experience severe weather, bandpass fall-offs (``frequency edge-channels''), and \ac{rfi}.\footnote{\ac{rfi} affects autocorrelation-based calibration steps and strong \ac{rfi} can also affect the crosscorrelations.}
In the unaveraged data, baseline-based problems are rare and the flagging should be done on a per-antenna basis.
There are some algorithms that can automatically identify some bad data, e.g., for \ac{rfi} \citep{2012A&A...539A..95O}, but to the authors' knowledge there is no software that is fully sufficient for \ac{vlbi} data.
The \ac{vlbi} data flagging therefore remains a subjective task.

Flags are usually applied at different data reduction stages. A priori flags based on telescope operator logs are made available to PIs.
Low \ac{sn} fringe non-detections render the underlying data flagged, but care must be taken here as the fringe-fitting segmentation time does not necessarily match the time ranges over which data issues are present. Partially affected data over \ac{vlbi} scan durations should be removed before fringe-fitting.
Finally, manifestations of bad data and calibration errors can easily be identified in image reconstructions as spurious features.
\Cref{fig:flag} shows two typical examples of bad data that should be flagged.

\reftitle{References}
\externalbibliography{yes}
\bibliography{main}

%


\end{document}